\documentclass[draftcls, onecolumn]{IEEEtran}

\usepackage{graphicx}
\usepackage{amsmath}
\usepackage{amssymb}
\usepackage{graphics}
\usepackage{latexsym}
\usepackage[dvips]{epsfig}
\usepackage[nospace]{cite}
\usepackage{subfigure}
\usepackage{setspace}
\usepackage{color}
\usepackage{tabularx}
\usepackage{url}

\usepackage{times}
\usepackage{textcomp}
\usepackage[left=0.625 in,top=0.82 in,right=0.625 in,bottom=1 in]{geometry}

\newtheorem{Definition}{Definition}
\newtheorem{Lemma}{Lemma}
\newtheorem{Corollary}[Lemma]{Corollary}
\newtheorem{Proposition}[Lemma]{Proposition}
\newtheorem{Theorem}{Theorem}

\newtheorem{Remark}{Remark}

\usepackage[ colorlinks = true,
             linkcolor = blue,
             urlcolor  = blue,
             citecolor = red,
             anchorcolor = green,
]{hyperref}

\allowdisplaybreaks[1]

\def\Pr{{\mathbb{P}}}
\def\E{{\mathrm E}}

\begin{document}
\title{Scaling Exponent and Moderate Deviations Asymptotics of Polar Codes for the AWGN Channel}

%\author{Silas~L.~Fong and Raymond~W.~Yeung,~\IEEEmembership{Fellow,~IEEE}% <-this % stops a space
%\thanks{Silas~L.~Fong is with the Department of Electronic Engineering, City University of Hong Kong, Kowloon, Hong Kong (e-mail: lhfong5@ie.cuhk.edu.hk).}% <-this % stops a space
%\thanks{Raymond~W.~Yeung is with the Institute of Network Coding and the Department
%of Information Engineering, The Chinese University of Hong Kong,
%N.T., Hong Kong (e-mail: whyeung@ie.cuhk.edu.hk).}}% <-this % stops a space
%%\thanks{Manuscript received April 19, 2005; revised January 11, 2007.}}

%\author{Silas~L.~Fong and Vincent~Y.~F.~Tan % <-this % stops a space
%\thanks{Silas~L.~Fong and Vincent~Y.~F.~Tan are with the Department of Electrical and Computer Engineering, National University of Singapore (NUS), Singapore (e-mail: \texttt{\{silas\_fong,vtan\}@nus.edu.sg}). Vincent~Y.~F.~Tan  is also with the Department of Mathematics, NUS. }
%%\thanks{$^*$Jing Yang is with the Department of Electrical Engineering at the University of Arkansas, AR, USA (email: \texttt{jingyang@uark.edu})}
%}

\author{Silas~L.~Fong,~\IEEEmembership{Member,~IEEE} and Vincent~Y.~F.~Tan,~\IEEEmembership{Senior Member,~IEEE}% <-this % stops a space
\thanks{S.~L.~Fong and V.~Y.~F.~Tan were supported by National University of Singapore (NUS) Young Investigator Award under Grant R-263-000-B37-133.}%
\thanks{S.~L.~Fong is with the Department of Electrical and Computer Engineering, NUS, Singapore 117583 (e-mail: \texttt{silas\_fong@nus.edu.sg}).}%
\thanks{V.~Y.~F.~Tan is with the Department of Electrical and Computer Engineering, NUS, Singapore 117583, and also with the Department of Mathematics, NUS, Singapore 119076 (e-mail: \texttt{vtan@nus.edu.sg}).}%
}
%\IEEEpubid{\parbox[c]{6 in}{\ \\ \\ 0000--0000/00\$00.00~\copyright~2016~IEEE. Personal use is permitted, but republication/redistribution requires IEEE permission.
% \\ \centering See \url{http://www.ieee.org/publications_standards/publications/rights/index.html} for more information.}}

\maketitle

\begin{abstract}
This paper investigates polar codes for the additive white Gaussian noise (AWGN) channel.
The scaling exponent~$\mu$ of polar codes for a memoryless channel~$q_{Y|X}$ with capacity $I(q_{Y|X})$ characterizes the closest gap between the capacity and non-asymptotic achievable rates in the following way: For a fixed $\varepsilon \in (0, 1)$, the gap between the capacity~$I(q_{Y|X})$ and the maximum non-asymptotic rate~$R_n^*$ achieved by a length-$n$ polar code with average error probability $\varepsilon$ scales as $n^{-1/\mu}$, i.e., $I(q_{Y|X})-R_n^* = \Theta(n^{-1/\mu})$.
 It is well known that the scaling exponent~$\mu$ for any binary-input memoryless channel (BMC) with $I(q_{Y|X})\in(0,1)$ is bounded above by $4.714$, which was shown by an explicit construction of polar codes. Our main result shows that $4.714$ remains to be a valid upper bound on the scaling exponent for the AWGN channel. Our proof technique involves the following two ideas: (i) The capacity of the AWGN channel can be achieved within a gap of $O(n^{-1/\mu}\sqrt{\log n})$ by using an input alphabet consisting of~$n$ constellations and restricting the input distribution to be uniform; (ii) The capacity of a multiple access channel (MAC) with an input alphabet consisting of~$n$ constellations can be achieved within a gap of $O(n^{-1/\mu}\log n)$ by using a superposition of $\log n$ binary-input polar codes.

In addition, we investigate the performance of polar codes in the moderate deviations regime where both the gap to capacity and the error probability vanish as~$n$ grows. An explicit construction of polar codes is proposed to obey a certain tradeoff between the gap to capacity and the decay rate of the error probability for the AWGN channel.
\end{abstract}


\IEEEpeerreviewmaketitle

\flushbottom

\section{Introduction} \label{Introduction}
\subsection{The Additive White Gaussian Noise Channel}
This paper investigates low-complexity codes over the classical additive white Gaussian noise (AWGN) channel~\cite[Ch.~9]{Cover06}, where a source wants to transmit information to a destination and each received symbol is the sum of the transmitted symbol and an independent Gaussian random variable. More specifically, if $X_k$ denotes the symbol transmitted by the source in the $k^{\text{th}}$ time slot, then the corresponding symbol received by the destination is
\begin{equation}
Y_k = X_k + Z_k \label{channelLaw}
\end{equation}
where $Z_k$ is the standard normal random variable. When the transmission lasts for~$n$ time slots, i.e., each transmitted codeword consists of~$n$ symbols, it is assumed that $Z_1, Z_2, \ldots, Z_n$ are independent and each transmitted codeword $x^n\triangleq(x_1, x_2, \ldots, x_n)$ must satisfy the peak power constraint
\begin{equation}
\frac{1}{n}\sum_{k=1}^n x_k^2 \le P
\end{equation}
where $P>0$ is a constant which denotes the permissible power.
%The capacity of the AWGN channel~\cite[Ch.~9]{Cover06} is well known to be
%\begin{equation}
%\mathrm{C}(P)\triangleq \frac{1}{2}\log(1+P), \label{defCP}
%\end{equation}
%which characterizes the limit of the maximum coding rate as~$n$ approaches infinity.
If we would like to transmit a uniformly distributed message $W\in \{1,2,\ldots, \lceil 2^{nR} \rceil\}$ across this channel, it was shown by Shannon \cite{Shannon48} that the limit of the maximum coding rate~$R$ as~$n$ approaches infinity (i.e., capacity) is
\begin{equation}
\mathrm{C}(P) \triangleq \frac{1}{2}\log(1+P). \label{defCP}
\end{equation}
\subsection{Polar Codes}
  Although the capacity of a memoryless channel was proved by Shannon~\cite{Shannon48} in 1948, low-complexity channel codes that achieve the capacity have not been found until Ar\i kan~\cite{Arikan} proposed to use polar codes with encoding and decoding complexities being $O(n\log n)$ for achieving the capacity of a binary-input memoryless symmetric channel (BMSC). This paper investigates the scaling exponent of polar codes~\cite{MHU15} for the AWGN channel, a ubiquitous channel model in wireless communications.

The scaling exponent~$\mu$ of polar codes for a memoryless channel~$q_{Y|X}$ with capacity
 \begin{align}
 I(q_{Y|X}) \triangleq \max_{p_X} I(X;Y) \label{defI}
 \end{align}
 characterizes the closest gap between the channel capacity and non-asymptotic achievable rates in the following way: For a fixed $\varepsilon \in (0, 1)$, the gap between the capacity~$I(q_{Y|X})$ and the maximum non-asymptotic rate~$R_n^*$ achieved by a length-$n$ polar code with average error probability $\varepsilon$ scales as $n^{-1/\mu}$, i.e., $I(q_{Y|X})-R_n^* = \Theta(n^{-1/\mu})$. It has been shown in \cite{HAU14, goldinBurshtein14, MHU15} that the scaling exponent~$\mu$ for any BMSC with $I(q_{Y|X})\in(0,1)$ lies between $3.579$ and $4.714$, where the upper bound $4.714$ was shown by an explicit construction of polar codes. Indeed, the upper bound~$4.714$ remains valid for any general binary-input memoryless channel (BMC) \cite[Lemma~4]{FongTan16polar}. It is well known that polar codes are capacity-achieving for BMCs \cite{STA09, SRDR12,HondaYamamoto13,MHU14Allerton}, and appropriately chosen ones are also capacity-achieving for the AWGN channel~\cite{AbbeBarron11}. In particular, for any $R<\mathrm{C}(P)$ and any $\beta<1/2$, polar codes operated at rate~$R$ can be constructed for the AWGN channel such that the decay rate of the error probability is $O(2^{-n^\beta})$~\cite{AbbeBarron11} and the encoding and decoding complexities are~$O(n\log n)$. However, the scaling exponent of polar codes for the AWGN channel has not been investigated yet.

In this paper, we construct polar codes for the AWGN channel and show that $4.714$ remains to be a valid upper bound on the scaling exponent. Our construction of polar codes involves the following two ideas: (i) By using an input alphabet consisting of~$n$ constellations and restricting the input distribution to be uniform as suggested in~\cite{AbbeBarron11}, we can achieve the capacity of the AWGN channel within a gap of $O(n^{-1/\mu}\sqrt{\log n})$; (ii) By using a superposition of $\log n$ binary-input polar codes\footnote{In this paper, $n$ is always a power of~$2$} as suggested in~\cite{AbbeTelatar12}, we can achieve the capacity of the corresponding multiple access channel (MAC) within a gap of $O(n^{-1/\mu}\log n)$ where the input alphabet of the MAC has~$n$ constellations (i.e., the size of the Cartesian product of the input alphabets corresponding to the $\log n$ input terminals is~$n$). The encoding and decoding complexities of our constructed polar codes are $O(n \log^2 n)$. On the other hand, the lower bound~$3.579$ holds trivially for the constructed polar codes because the polar codes are constructed by superposing $\log n$ binary-input polar codes whose scaling exponents are bounded below by~$3.579$ \cite{HAU14}.

In addition, Mondelli et al.\ \cite[Sec.~IV]{MHU15} provided an explicit construction of polar codes for any BMSC which obey a certain tradeoff between the gap to capacity and the decay rate of the error probability. More specifically, if the gap to capacity is set to vanish at a rate of $\Theta\left(n^{-\frac{1-\gamma}{\mu}}\right)$ for some $\gamma\in\left(\frac{1}{1+\mu}, 1\right)$, then a length-$n$ polar code can be constructed such that the error probability is~$O\left(n \cdot 2^{-n^{\gamma h_2^{-1}\left(\frac{\gamma\mu + \gamma -1}{\gamma \mu}\right)}}\right)$ where $h_2:[0, 1/2]\rightarrow[0,1]$ denotes the binary entropy function. This tradeoff was developed under the moderate deviations regime~\cite{altug14b} where both the gap to capacity and the error probability vanish as~$n$ grows. For the AWGN channel, we develop a similar tradeoff under the moderate deviations regime by using our constructed polar codes described above.

\subsection{Paper Outline}
This paper is organized as follows. The notation used in this paper is described in the next subsection. Section~\ref{sectionDefinitionPTP} presents the background of this work, which includes existing polarization results for the BMC which are used in this work. Sections~\ref{subsec2D} to~\ref{subsec2F} state the formulation of the binary-input MAC and present new polarization results for the binary-input MAC. Sections~\ref{subsec2G} to~\ref{subsec2H} state the formulation of the AWGN channel and present new polarization results for the AWGN channel. Section~\ref{sectionMainResult} establishes the definition of the scaling exponent for the AWGN channel and establishes the main result --- 4.714 is an upper bound on the scaling exponent of polar codes for the AWGN channel. Section~\ref{sectionModerateDeviation} presents an explicit construction of polar codes for the AWGN channel which obey a certain tradeoff between the gap to capacity and the decay rate of the error probability under the moderate deviations regime. Concluding remarks are provided in Section~\ref{sectionConclusion}.
%%%

 \subsection{Notation} \label{sectionNotation}
%We use $\Pr\{\mathcal{E}\}$ to represent the probability of an
%event~$\mathcal{E}$, and
The set of natural numbers, real numbers and non-negative real numbers are denoted by $\mathbb{N}$, $\mathbb{R}$ and $\mathbb{R}_+$ respectively. For any sets $\mathcal{A}$ and $\mathcal{B}$ and any mapping $f:\mathcal{A}\rightarrow \mathcal{B}$, we let $f^{-1}(\mathcal{D})$ denote the set $\{a\in\mathcal{A}\,|f(a)\in\mathcal{D}\}$ for any $\mathcal{D}\subseteq\mathcal{B}$. We let $\boldsymbol{1}\{\mathcal{E}\}$ be the indicator function of the set $\mathcal{E}$.
An arbitrary (discrete or continuous) random variable is denoted by an upper-case letter (e.g., $X$), and the realization and the alphabet of the random variable are denoted by the corresponding lower-case letter (e.g., $x$) and calligraphic letter (e.g., $\mathcal{X}$) respectively.
%We use $X_{(1, \ldots, N)}$ to denote a random tuple $(X_1, X_2, \ldots, X_N)$, where each element $X_k$ is chosen from some alphabet $\mathcal{X}_k$ for $i=1, 2, \ldots, N$.
We use $X^n$ to denote the random tuple $(X_1, X_2, \ldots, X_n)$ where each $X_k$ has the same alphabet $\mathcal{X}$. We will take all logarithms to base $2$ throughout this paper. %The logarithmic functions to base~$2$ and base~$e$ are denoted by $\log$ and $\ln$ respectively.
%, and we use $\{X^n\}_{k=1}^n$ to denote the sequence $X_1, X_2, \ldots, X_n$. %, where the components $X_k$ have the same alphabet~$\mathcal{X}$.

The following notations are used for any arbitrary random variables~$X$ and~$Y$ and any real-valued function $g$ with domain $\mathcal{X}$. We let $p_{Y|X}$ and $p_{X,Y}=p_Xp_{Y|X}$ denote the conditional probability distribution of $Y$ given $X$ and the probability distribution of $(X,Y)$ respectively.
% (can be both discrete, both continuous or one discrete and one continuous)
We let $p_{X,Y}(x,y)$ and $p_{Y|X}(y|x)$ be the evaluations of $p_{X,Y}$ and $p_{Y|X}$ respectively at $(X,Y)=(x,y)$. %To avoid confusion, we do not write $\Pr\{X=x, Y=y\}$ to represent $p_{X,Y}(x,y)$ unless $X$ and $Y$ are both discrete.
%We let $p_{g(X)}$ denote the probability distribution of $g(X)$ when $X$ is distributed according to $p_X$, and
To make the dependence on the distribution explicit, we let $\Pr_{p_X}\{ g(X)\in\mathcal{A}\}$ denote $\int_{\mathcal{X}} p_X(x)\mathbf{1}\{g(x)\in\mathcal{A}\}\, \mathrm{d}x$ for any set $\mathcal{A}\subseteq \mathbb{R}$.
The expectation of~$g(X)$ is denoted as
$\E_{p_X}[g(X)]$. For any $(X,Y, Z)$ distributed according to some $p_{X,Y, Z}$, the entropy of $X$ and the conditional mutual information between $X$ and $Y$ given~$Z$ are denoted by $H_{p_X}(X)$ and $I_{p_{X,Y,Z}}(X;Y|Z)$ respectively. For simplicity, we sometimes omit the subscript of a notation if it causes no confusion.
 %The total variation distance between $p_X$ and $q_X$ is denoted by \[\|p_X-q_X\|\triangleq \frac{1}{2}\sum_{x\in \mathcal{X}}|p_X(x)-q_X(x)|.\]%, where we again make the dependence on the underlying distribution $p_X$ explicit.
% We let $p_Xp_{Y|X}$ denote the joint distribution of $(X,Y)$, i.e., $p_Xp_{Y|X}(x,y)=p_X(x)p_{Y|X}(y|x)$ for all $x$ and $y$.
%The total variation distance between $p_X$ and $q_X$ is denoted by \[\|p_X-q_X\|\triangleq \frac{1}{2}\sum_{x\in \mathcal{X}}|p_X(x)-q_X(x)|.\]
 The relative entropy between $p_X$ and $q_X$ is denoted by
 \begin{equation}
 D(p_X\|q_X)\triangleq \int_{\mathcal{X}} p_X(x)\log\left(\frac{p_X(x)}{q_X(x)}\right) \mathrm{d}x.
 \end{equation}
  The $2$-Wasserstein distance between $p_X$ and $p_Y$ is denoted by
 \begin{equation}
W_2(p_X, p_Y)\triangleq \inf_{\substack{s_{X,Y}: \\ s_X=p_X,\\ s_Y=p_Y}}\sqrt{\int_{\mathcal{X}}\int_{\mathcal{Y}} s_{X,Y}(x,y)(x-y)^2 \mathrm{d}y  \mathrm{d}x}\,. \label{defW2Distance}
 \end{equation}
 %The $\chi^2$-distance between $p_X$ and $q_X$ is denoted by
% \begin{equation}
% \chi^2(p_X\|q_X)\triangleq \int_{\mathcal{X}} \frac{(p_X(x)-q_X(x))^2}{q_X(x)}\,\mathrm{d}x. \label{chiSquareDiv}
%  \end{equation}
% It is well known \cite[Th.~5]{GibbsSu2002} that
% \begin{equation}
% D(p_X\|q_X) \le  \chi^2(p_X\|q_X). \label{divInequality}
% \end{equation}
 We let $\mathcal{N}(\,\cdot\, ;\mu,\sigma^2): \mathbb{R}\rightarrow [0,\infty)$ denote the probability density function of a Gaussian random variable whose mean and variance are $\mu$ and $\sigma^2$ respectively, i.e., \begin{equation}
\mathcal{N}(z;\mu,\sigma^2)\triangleq\frac{1}{\sqrt{2\pi \sigma^2}}e^{-\frac{(z-\mu)^2}{2\sigma^2} }. \label{eqnNormalDist}
\end{equation}
%We will use the convention that $0\log 0=0\log \frac{0}{0}=0$ throughout this paper.

\section{Background: Point-to-Point Channels and Existing Polarization Results}
\label{sectionDefinitionPTP}
In this section, we will review important polarization results related to the scaling exponent of polar codes for binary-input memoryless channels (BMCs).
\subsection{Point-to-Point Memoryless Channels} \label{subsecPTP}
Consider a point-to-point channel which consists of one source and one destination, denoted by $\mathrm{s}$ and $\mathrm{d}$ respectively. Suppose node~$\mathrm{s}$ transmits information to node~$\mathrm{d}$ in $n$ time slots. Before any transmission begins, node~$\mathrm{s}$ chooses message
$W$ destined for node~$\mathrm{d}$, where $W$ is uniformly distributed over the alphabet
\begin{equation}
\mathcal{W}\triangleq\{1, 2, \ldots, M\} \label{defMessageSet}
\end{equation}
which consists of~$M$ elements.
For each $k\in \{1, 2, \ldots, n\}$, node~$\mathrm{s}$ transmits $X_k\in \mathcal{X}$ based on~$W$ and node~$\mathrm{d}$ receives $Y_k\in \mathcal{Y}$ in time slot~$k$ where $\mathcal{X}$ and $\mathcal{Y}$ denote respectively the input and output alphabets of the channel.
   After~$n$ time slots, node~$\mathrm{d}$ declares~$\hat W$ to be the transmitted~$W$ based on~$Y^n$. Formally, we define a length-$n$ code as follows.
   \medskip
\begin{Definition} \label{defCode}
An {\em $(n, M)$-code} consists of the following:
\begin{enumerate}
\item A message set
$
\mathcal{W}
$
as defined in~\eqref{defMessageSet}. Message $W$ is uniform on $\mathcal{W}$.

\item An encoding function
$
f_k : \mathcal{W}\rightarrow \mathcal{X}
$
for each $k\in\{1, 2, \ldots, n\}$,
where~$f_k$ is used by node~$\mathrm{s}$ for encoding $X_k$ such that
$
X_k=f_k(W)$.
%\begin{equation}
%\Pr\left\{\sum_{\ell=1}^k X_\ell^2 \le \sum_{\ell=1}^k E_\ell\right\} = 1. \label{powerConstraint}
%\end{equation}
\item A decoding function
$
\varphi :
\mathcal{Y}^{n} \rightarrow \mathcal{W}
$
%where $\varphi$ is the decoding function
 used by node~$\mathrm{d}$ for producing the message estimate
$
\hat W = \varphi(Y^{n})$.
\end{enumerate}
%If the sequence of encoding functions $f_k$ satisfies the EH constraints~\eqref{eqn:eh}, the code is also called an {\em $(n, M)$-EH code}.
\end{Definition}
\medskip

\begin{Definition}\label{defChannel}
The {\em point-to-point memoryless channel} is characterized by an input alphabet~$\mathcal{X}$, an output alphabet~$\mathcal{Y}$ and a conditional distribution $q_{Y|X}$ such that the following holds for any $(n, M)$-code: For each $k\in\{1, 2, \ldots, n\}$,
$
p_{W, X^n, Y^n}
 = p_{W, X^n}\prod_{k=1}^n p_{Y_k|X_k} %\label{memorylessStatement*}
$
where
$
p_{Y_k|X_k}(y_k|x_k) = q_{Y|X}(y_k|x_k) $
for all $x_k\in \mathcal{X}$ and $y_k\in \mathcal{Y}$.
%Since $p_{Y_k|X_k}$ does not depend on~$i$ by \eqref{defChannelInDefinition*}, the channel is stationary.
\end{Definition}
%\medskip
%\begin{Definition}
%The binary-input channel $q_{Y|X}$ is said to be \textit{symmetric} if there exists a permutation $\pi$ of the output alphabet $\mathcal{Y}$ such that (i) $\pi^{-1}=\pi$ and (ii) $q_{Y|X}(y|1)=q_{Y|X}(\pi(y)|0)$ for all $y\in \mathcal{Y}$. Otherwise, the channel is said to be \textit{asymmetric}.
%\end{Definition}
%\medskip

 For any $(n, M)$-code defined on the point-to-point memoryless channel, let $p_{W, X^n, Y^n, \hat W}$ be the joint distribution induced by the code. By Definitions~\ref{defCode} and~\ref{defChannel}, we can factorize $p_{W, X^n, Y^n, \hat W}$ as
\begin{align}
 p_{W, X^n, Y^n, \hat W}
%&\stackrel{\text{(a)}}{=} p_{W,E^n, X^n, Y^n}p_{\hat W |Y^n}  \notag\\
%&= p_{W} \left(\prod_{k=1}^n p_{E_k X_k,Y_k|W, E^{k-1},X^{k-1}, Y^{k-1}}\right)p_{\hat W |Y^n} \notag\\
%&= p_{W} \left(\prod_{k=1}^n p_{E_k|W, E^{k-1},X^{k-1}, Y^{k-1}}p_{X_k,Y_k|W, E^k,X^{k-1}, Y^{k-1}}\right)p_{\hat W |Y^n} \notag\\
%&\stackrel{\eqref{assumption(i)}}{=} p_{W} \left(\prod_{k=1}^n p_{E_k}p_{X_k,Y_k|W, E^k,X^{k-1}, Y^{k-1}}\right)p_{\hat W |Y^n} \notag\\
%&= p_W \left(\prod_{k=1}^n  p_{E_k} p_{X_k|W, E^k, X^{k-1}, Y^{k-1}}p_{Y_k|W, E^k, X^k, Y^{k-1}} \right)p_{\hat W |Y^n} \notag\\
%& \stackrel{\text{(b)}}{=} p_W \left(\prod_{k=1}^n p_{E_k}p_{X_k|W, E^k} p_{Y_k|W, E^k, X^k, Y^{k-1}} \right)p_{\hat W |Y^n}\notag\\
% \stackrel{\text{(c)}}{=}
=p_W p_{X^n|W}\bigg(\prod_{k=1}^n p_{Y_k|X_k}\bigg)p_{\hat W |Y^n}. \label{memorylessStatement}
\end{align}

\subsection{Polarization for Binary-Input Memoryless Channels} \label{subsec2B}
\begin{Definition}\label{defBMC}
A point-to-point memoryless channel characterized by $q_{Y|X}$ is called a \emph{binary-input memoryless channel (BMC)} if $\mathcal{X}=\{0,1\}$.
\end{Definition}

\medskip
We follow the formulation of polar coding in~\cite{HondaYamamoto13}. Consider any BMC characterized by $q_{Y|X}$.
%  \begin{equation}
%(Z_{p_{U, X, Y}}(U|Y))^2 \le H_{p_{U,X,Y}}(U|Y). \label{bhattacharyyaEntropy}
%  \end{equation}
  Let $p_X$ be the probability distribution of a Bernoulli random variable~$X$, and let $p_{X^n}$ be the distribution of~$n$ independent copies of~$X\sim p_X$, i.e., $p_{X^n}(x^n) = \prod_{k=1}^n p_{X}(x_k)$ for all $x^n\in \mathcal{X}^n$. For each $n=2^m$ where $m\in \mathbb{N}$, the polarization mapping of a length-$n$ polar code is given by
  \begin{equation}
  G_n \triangleq \bigg[\,\begin{matrix} 1& 0  \\  1 & 1 \end{matrix}\,\bigg]^{\otimes m} = G_n^{-1} \label{defGn}
  \end{equation}
  where $\otimes$ denotes the Kronecker power. Define $p_{U^n|X^n}$ such that
  \begin{equation}
[U_1\ U_2\ \ldots \ U_n]  = [X_1\ X_2\ \ldots \ X_n] G_n \label{defpUgivenX}
  \end{equation}
  where the addition and product operations are performed over GF(2),
define
  \begin{equation*}
  p_{Y_k|X_k}(y_k|x_k) \triangleq q_{Y|X}(y_k|x_k)
  \end{equation*}
  for each $k\in\{1, 2, \ldots, n\}$ and each $(x_k, y_k)\in \mathcal{X}\times \mathcal{Y}$ where $q_{Y|X}$ characterizes the BMC (cf.\ \eqref{defChannel}),
  and define
  \begin{equation}
  p_{U^n, X^n, Y^n}\triangleq p_{X^n} p_{U^n|X^n}\prod_{k=1}^n p_{Y_k|X_k}. \label{defP}
  \end{equation}
  In addition, for each $k\in\{1, 2, \ldots, n\}$, define the Bhattacharyya parameter associated with time~$k$ as
\begin{align}
&Z^{[p_X; q_{Y|X}]}(U_k|U^{k-1}, Y^n) \notag\\*
& \quad\triangleq  2\sum_{u^{k-1}\in\mathcal{U}^{k-1}}\int_{\mathcal{Y}^n} p_{U^{k-1}, Y^n}(u^{k-1}, y^n)\sqrt{p_{U_k|U^{k-1}, Y^n}(0|u^{k-1}, y^n)p_{U_k|U^{k-1}, Y^n}(1|u^{k-1}, y^n)}\mathrm{d}y^n\label{defBhattacharyya*}\\*
& \quad= 2\sum_{u^{k-1}\in\mathcal{U}^{k-1}}\int_{\mathcal{Y}^n}\sqrt{p_{U_k,U^{k-1}, Y^n}(0, u^{k-1}, y^n)p_{U_k,U^{k-1}, Y^n}(1,u^{k-1}, y^n)}\mathrm{d}y^n, \label{defBhattacharyya}
  \end{align}
  where the distributions in~\eqref{defBhattacharyya*} and~\eqref{defBhattacharyya} are marginal distributions of $p_{U^n, X^n, Y^n}$ defined in~\eqref{defP}.
%  It is well known that \cite[Proposition~2]{Arikan:10ISIT}
 % where
%  \begin{equation}
%  p_{Y^n|X^n}(y^n|x^n)=\prod_{k=1}^n q_{Y|X}(y_k|x_k) \label{defPYnGivenXn}
%  \end{equation}
The following result is based on~\cite[Sec.~III]{MHU15} and has been used in \cite{FongTan16polar} to show that $4.714$ is an upper bound on the scaling exponent for any BMC. To simplify notation, let
\begin{equation}
\beta\triangleq 4.714
\end{equation}
 in the rest of this paper.
%The proof combines key ideas in~\cite{MHU15} and~\cite{HondaYamamoto13}, and is relegated to Appendix~\ref{appendixA}.
\medskip
\begin{Lemma}[{\cite[Sec.~III]{MHU15},\cite[Lemma~2]{FongTan16polar}}]\label{lemmaPolar}
 There exists a universal constant $t>0$ such that the following holds. Fix any BMC characterized by $q_{Y|X}$ and any $p_X$. Then for any $m\in\mathbb{N}$ and $n\triangleq 2^m$, we have\footnote{This lemma remains to hold if the quantities $\frac{1}{n^4}$ are replaced by $\frac{1}{n^\nu}$ for any $\nu>0$. The main result of this paper continues to hold if the quantities $\frac{1}{n^4}$ in this lemma are replaced by $\frac{1}{n^\nu}$ for any $\nu > 2$.}
 \begin{align}
 \frac{1}{n}\left|\left\{ k\in\{1, 2, \ldots, n\}\left|\parbox[c]{1.85 in}{$ Z^{[p_X; q_{Y|X}]}(U_k|U^{k-1}, Y^n) \le \frac{1}{n^4}$} \!\! \right.  \right\}\right| \ge I_{p_X q_{Y|X}}(X; Y) - \frac{t}{n^{1/\beta}}. \label{st1InLemmaPolar}
 \end{align}
%and
%  \begin{align*}
%& \frac{1}{n}\!\left|\left\{ k\in\{1, 2, \ldots, n\}\!\left| \parbox[c]{2.18 in}{$ Z_{p_{U^n, X^n, Y^n}}(U_k|U^{k-1}, Y^n) \ge 1-  \frac{1}{n^4}, \\Z_{p_{U^n, X^n, Y^n}}(U_k|U^{k-1}) \le \frac{1}{n^4}$} \!\!\right.  \right\}\right| \notag\\
% &\quad\ge 1- I_{p_X q_{Y|X}}(X; Y) - \frac{t_2}{n^{1/\beta}}. % \label{st2InLemmaPolar}
% \end{align*}
\end{Lemma}

\section{Problem Formulation of Binary-Input MACs and New Polarization Results}
\label{sectionDefinitionMAC}
Polar codes have been proposed and investigated for achieving any rate tuple inside the capacity region of a binary-input multiple access channel (MAC) \cite{AbbeTelatar12, MELK16}.
The goal of this section is to use the polar codes proposed in~\cite{AbbeTelatar12} to achieve the symmetric sum-capacity of a binary-input MAC.
%generalize the scaling result for BMCs in Proposition~\ref{propositionBMCscaling} to binary-input multiple access channels (MACs).
\subsection{Binary-Input Multiple Access Channels} \label{subsec2D}
%We would like to generalize polar codes for BMCs described in Definition~\ref{defPolarCodeWithFrozenBits} for binary-input multiple access channels. To this end, we first formulate a multiple access channel (MAC) as follows.
Consider a MAC~\cite[Sec.~15.3]{Cover06} which consists of $N$ sources and one destination. Let $\mathcal{I}\triangleq\{1, 2, \ldots, N\}$ be the index set of the $N$ sources and let $\mathrm{d}$ denote the destination. Suppose the sources transmit information to node~$\mathrm{d}$ in~$n$ time slots. Before any transmission begins, node~$i$ chooses message
$W_i$ destined for node~$\mathrm{d}$ for each $i\in\mathcal{I}$, where $W_i$ is uniformly distributed over
\begin{equation}
\mathcal{W}_i\triangleq\{1, 2, \ldots, M_i\} \label{defMessageSetMAC}
\end{equation}
which consists of~$M_i$ elements.
For each $k\in \{1, 2, \ldots, n\}$, node~$i$ transmits $X_{i,k}\in \mathcal{X}_i$ based on~$W_i$ for each $i\in\mathcal{I}$ and node~$\mathrm{d}$ receives $Y_k\in \mathcal{Y}$ in time slot~$k$ where $\mathcal{X}_i$ denotes the input alphabet for node~$i$ and $\mathcal{Y}$ denotes the output alphabet. After~$n$ time slots, node~$\mathrm{d}$ declares~$\hat W_i$ to be the transmitted~$W_i$ based on~$Y^n$ for each $i\in\mathcal{I}$.

To simplify notation, we use the following convention for any $T\subseteq\mathcal{I}$. For any random tuple $(X_1, X_2, \ldots, X_N)$, we let
$
X_T \triangleq (X_i: i\in T)
$
be the corresponding subtuple, whose realization and alphabet are denoted by $x_T$ and $\mathcal{X}_T$ respectively. Similarly, for each $k\in\{1, 2, \ldots, n\}$ and each random tuple $(X_{1, k}, X_{2,k}, \ldots, X_{N,k})\in \mathcal{X}_\mathcal{I}$, we let
$
X_{T,k} \triangleq (X_{i,k}: i\in T)
$
denote the corresponding random subtuple, and let $x_{T,k}$ and $\mathcal{X}_{T,k}$ denote respectively the realization and the alphabet of $X_{T,k}$.
 Formally, we define a length-$n$ code for the binary-input MAC as follows.
   \medskip

   \begin{Definition} \label{defCodeMAC}
An {\em $(n, M_\mathcal{I})$-code}, where $M_\mathcal{I}\triangleq(M_1, M_2, \ldots, M_N)$, consists of the following:
\begin{enumerate}
\item A message set
$
\mathcal{W}_i
$
for each $i\in\mathcal{I}$ as defined in~\eqref{defMessageSetMAC}, where message $W_i$ is uniform on $\mathcal{W}_i$.

\item An encoding function
$
f_{i,k}^{\text{MAC}} : \mathcal{W}_i\rightarrow \mathcal{X}_i
$
for each $i\in\mathcal{I}$ and each $k\in\{1, 2, \ldots, n\}$, where~$f_{i,k}^{\text{MAC}}$ is used by node~$i$ for encoding $X_{i,k}$ such that
$
X_{i,k}=f_{i,k}^{\text{MAC}}(W_i)$.
%\begin{equation}
%\Pr\left\{\sum_{\ell=1}^k X_\ell^2 \le \sum_{\ell=1}^k E_\ell\right\} = 1. \label{powerConstraint}
%\end{equation}
\item A decoding function
$
\varphi^{\text{MAC}} :
\mathcal{Y}^{n} \rightarrow \mathcal{W}_\mathcal{I}
$
%where $\varphi$ is the decoding function
 used by node~$\mathrm{d}$ for producing the message estimate
$
\hat W_\mathcal{I} = \varphi^{\text{MAC}}(Y^{n})$.
\end{enumerate}
%If the sequence of encoding functions $f_k$ satisfies the EH constraints~\eqref{eqn:eh}, the code is also called an {\em $(n, M)$-EH code}.
\end{Definition}
\medskip

\begin{Definition}\label{defChannelMAC}
The {\em multiple access channel (MAC)} is characterized by~$N$ input alphabets specified by~$\mathcal{X}_\mathcal{I}$, an output alphabet specified by~$\mathcal{Y}$ and a conditional distribution $q_{Y|X_\mathcal{I}}$ such that the following holds for any $(n, M_\mathcal{I})$-code: For each $k\in\{1, 2, \ldots, n\}$,
\begin{align}
p_{W_\mathcal{I}, X_\mathcal{I}^n, Y^n}
 = \bigg(\prod_{i\in\mathcal{I}} p_{W_i, X_i^n}\bigg)\prod_{k=1}^n p_{Y_k|X_{\mathcal{I},k}} %\label{memorylessStatement*}
\end{align}
where
$
p_{Y_k|X_{\mathcal{I},k}}(y_k|x_{\mathcal{I},k}) = q_{Y|X_\mathcal{I}}(y_k|x_{\mathcal{I},k}) %\label{defChannelInDefinition*}
$
for all $x_{\mathcal{I},k}\in \mathcal{X}_\mathcal{I}$ and $y_k\in \mathcal{Y}$.
%Since $p_{Y_k|X_k}$ does not depend on~$i$ by \eqref{defChannelInDefinition*}, the channel is stationary.
\end{Definition}

\subsection{Polarization for Binary-Input MACs} \label{subsec2E}
\begin{Definition}\label{defChannelBinaryMAC}
A MAC characterized by $q_{Y|X_\mathcal{I}}$ is called a \textit{binary-input MAC} if $\mathcal{X}_\mathcal{I} = \{0,1\}^N$.
\end{Definition}

\medskip
Consider any binary-input MAC characterized by $q_{Y|X_\mathcal{I}}$.
%  \begin{equation}
%(Z_{p_{U, X, Y}}(U|Y))^2 \le H_{p_{U,X,Y}}(U|Y). \label{bhattacharyyaEntropy}
%  \end{equation}
For each $i\in\mathcal{I}$, let $p_{X_i}$ be the probability distribution of a Bernoulli random variable~$X_i$, and let $p_{X_i^n}$ be the distribution of~$n$ independent copies of~$X_i\sim p_{X_i}$, i.e., $p_{X_i^n}(x_i^n) = \prod_{k=1}^n p_{X_i}(x_{i,k})$ for all $x_i^n\in \mathcal{X}_i^n$.
%For each $n=2^m$ where $m\in \mathbb{N}$, the polarization mapping of a length-$n$ polar code is given by
%  \begin{equation}
%  G_n \triangleq \bigg[\begin{matrix} 1& 0  \\  1 & 1 \end{matrix}\bigg]^{\otimes m} = G_n^{-1} \label{defGn}
%  \end{equation}
%  where $\otimes$ denotes the Kronecker power.
  Recall the polarization mapping $G_n$ defined in~\eqref{defGn}.
 For each $i\in\mathcal{I}$, define $p_{U_i^n|X_i^n}$ such that
  \begin{equation}
[U_{i,1}\ U_{i,2}\ \ldots \ U_{i,n}]  = [X_{i,1}\ X_{i,2}\ \ldots \ X_{i,n}] G_n \label{defpUgivenX}
  \end{equation}
   where the addition and product operations are performed over GF(2),
%define
%  \begin{equation*}
%  p_{Y_k|X_{\mathcal{I},k}}(y_k|x_{\mathcal{I},k}) \triangleq q_{Y|X_\mathcal{I}}(y_k|x_{\mathcal{I},k})
%  \end{equation*}
%  for each $k\in\{1, 2, \ldots, n\}$ and each $(x_{\mathcal{I},k}, y_k)\in \mathcal{X}_\mathcal{I}\times \mathcal{Y}$,
%  and
and define
  \begin{equation}
  p_{U_\mathcal{I}^n, X_\mathcal{I}^n, Y^n}\triangleq \bigg(\prod_{i\in\mathcal{I}}p_{X_i^n} p_{U_i^n|X_i^n}\bigg)\prod_{k=1}^n p_{Y_k|X_{\mathcal{I},k}}. \label{defPmac}
  \end{equation}
  In addition, for each $i\in\mathcal{I}$ and each $k\in\{1, 2, \ldots, n\}$, define $[i-1] \triangleq \{1, 2, \ldots, i-1\}$ and define the Bhattacharyya parameter associated with node~$i$ and time~$k$ as
\begin{align}
&Z^{[p_{X_\mathcal{I}}; q_{Y|X_\mathcal{I}}]}(U_{i,k}|U_i^{k-1}, X_{[i-1]}^n, Y^n) \notag\\*
%& \quad\triangleq  2\sum_{u^{k-1}\in\mathcal{U}^{k-1}}\int_{\mathcal{Y}^n} p_{U^{k-1}, Y^n}(u^{k-1}, y^n)\sqrt{p_{U_k|U^{k-1}, Y^n}(0|u^{k-1}, y^n)p_{U_k|U^{k-1}, Y^n}(1|u^{k-1}, y^n)}\mathrm{d}y^n\label{defBhattacharyya*MAC}\\*
& \quad\triangleq 2\sum_{u_i^{k-1}\in\mathcal{U}_i^{k-1}}\sum_{x_{[i-1]}^{n}\in \{0,1\}^{(i-1)n}}\int_{\mathcal{Y}^n}\sqrt{p_{U_{i,k},U_i^{k-1}, X_{[i-1]}^n, Y^n}(0, u_i^{k-1}, x_{[i-1]}^n, y^n)p_{U_{i,k},U_i^{k-1},X_{[i-1]}^n, Y^n}(1,u_i^{k-1},x_{[i-1]}^n, y^n)}\mathrm{d}y^n, \label{defBhattacharyyaMAC}
\end{align}
  where the distributions in~\eqref{defBhattacharyyaMAC} are marginal distributions of $p_{U_\mathcal{I}^n, X_\mathcal{I}^n, Y^n}$ defined in~\eqref{defPmac}.
%  It is well known that \cite[Proposition~2]{Arikan:10ISIT}
 % where
%  \begin{equation}
%  p_{Y^n|X^n}(y^n|x^n)=\prod_{k=1}^n q_{Y|X}(y_k|x_k) \label{defPYnGivenXn}
%  \end{equation}
The following lemma is a direct consequence of Lemma~\ref{lemmaPolar}.
%The proof combines key ideas in~\cite{MHU15} and~\cite{HondaYamamoto13}, and is relegated to Appendix~\ref{appendixA}.
\medskip
\begin{Lemma}\label{lemmaPolarMAC}
%Let $\beta=4.714$.
 There exists a universal constant $t>0$ such that the following holds. Fix any binary-input MAC characterized by $q_{Y|X_\mathcal{I}}$ and any $p_{X_\mathcal{I}}$. Then for any $m\in\mathbb{N}$ and $n\triangleq 2^m$, we have\footnote{This lemma remains to hold if the quantities $\frac{1}{n^4}$ are replaced by $\frac{1}{n^\nu}$ for any $\nu>0$. The main result of this paper continues to hold if the quantities $\frac{1}{n^4}$ in this lemma are replaced by $\frac{1}{n^\nu}$ for any $\nu > 2$.}
 \begin{align}
 \frac{1}{n}\left|\left\{ k\in\{1, 2, \ldots, n\}\left|\parbox[c]{2.45 in}{$ Z^{[p_{X_\mathcal{I}}; q_{Y|X_\mathcal{I}}]}(U_{i,k}|U_i^{k-1}, X_{[i-1]}^n, Y^n)  \le \frac{1}{n^4}$}  \right.  \right\}\right| \ge I_{p_{X_\mathcal{I}} q_{Y|X_\mathcal{I}}}(X_i; X_{[i-1]},Y) - \frac{t}{n^{1/\beta}} \label{st1InLemmaPolarMAC}
 \end{align}
 for each $i\in\mathcal{I}$.
% \begin{align}
% \frac{1}{n}\left|\left\{ k\in\{1, 2, \ldots, n\}\left|\parbox[c]{1.9 in}{$ Z^{[p_X; q_{Y|X}]}(U_k|U^{k-1}, Y^n) \le \frac{1}{n^4}, \vspace{0.04 in} \\
% Z^{[p_X; q_{Y|X}]}(U_k|U^{k-1}) \ge 1-\frac{1}{n^4}$}  \right.  \right\}\right| \ge I_{p_{X_\mathcal{I}} q_{Y|X_\mathcal{I}}}(X_i; X_{[i-1]},Y) - \frac{t}{n^{1/\beta}}. \label{st1InLemmaPolar}
% \end{align}
%and
%  \begin{align*}
%& \frac{1}{n}\!\left|\left\{ k\in\{1, 2, \ldots, n\}\!\left| \parbox[c]{2.18 in}{$ Z_{p_{U^n, X^n, Y^n}}(U_k|U^{k-1}, Y^n) \ge 1-  \frac{1}{n^4}, \\Z_{p_{U^n, X^n, Y^n}}(U_k|U^{k-1}) \le \frac{1}{n^4}$} \!\!\right.  \right\}\right| \notag\\
% &\quad\ge 1- I_{p_X q_{Y|X}}(X; Y) - \frac{t_2}{n^{1/\beta}}. % \label{st2InLemmaPolar}
% \end{align*}
\end{Lemma}
\begin{IEEEproof}
Fix any $i\in\mathcal{I}$. Construct $p_{X_{[i-1]}, Y|X_i}$ by marginalizing $p_{X_\mathcal{I}} q_{Y|X_\mathcal{I}}$ and view $p_{X_{[i-1]}, Y|X_i}$ as the conditional distribution that characterizes a BMC. The lemma then follows directly from Lemma~\ref{lemmaPolar}.
\end{IEEEproof}
%\medskip
%\begin{Remark}
%The bound in~\eqref{st1InLemmaPolarMAC} in Lemma~\ref{lemmaPolarMAC} tells us that the fraction of good synthesized channels associated with node~$i$ is close to the mutual information $I(X_i; X_{[i-1]}, Y)$ in terms of their Bhattachryya parameters. Furthermore the notions of ``good" and ``close to $I(X_i; X_{[i-1]}, Y)$" are quantified precisely as functions of~$n$. %These quantifications of the rates of convergence allow us to establish a meaningful bound on the scaling exponent.
%\end{Remark}

\subsection{Polar Codes That Achieve the Symmetric Sum-Capacity of a Binary-Input MAC} \label{subsec2F}
Throughout this paper, let $p_{X_i}^*$ denote the uniform distribution on $\{0,1\}$ for each $i\in\mathcal{I}$ and define $p_{X_\mathcal{I}}^*\triangleq\prod_{i\in\mathcal{I}}p_{X_i}^*$, i.e.,
\begin{align}
p_{X_\mathcal{I}}^*(x_\mathcal{I})=\frac{1}{2^N} \label{defPI*}
\end{align}
for any $x_\mathcal{I}\in \{0,1\}^N$.
\medskip
\begin{Definition}\label{defSymmetricCapacityMAC}
For a binary-input MAC characterized by $q_{Y|X_\mathcal{I}}$, the symmetric sum-capacity is defined to be \linebreak$C_{\text{sum}} \triangleq I_{p_{X_\mathcal{I}}^*q_{Y|X_\mathcal{I}}}(X_\mathcal{I};Y)$.
\end{Definition}
\medskip

The following definition summarizes the polar codes for the binary-input MAC proposed in~\cite[Sec.~IV]{AbbeTelatar12}.
%a generalization of Definition~\ref{defPolarCodeWithFrozenBits}.
%\medskip
\begin{Definition}[{\cite[Sec.~IV]{AbbeTelatar12}}] \label{defPolarCodeWithFrozenBitsMAC}
 Fix an $n=2^m$ where $m\in \mathbb{N}$. For each $i\in\mathcal{I}$, let $\mathcal{J}_i$ be a subset of $\{1, 2, \ldots, n\}$, define $\mathcal{J}_i^c \triangleq \{1, 2, \ldots, n\}\setminus\mathcal{J}_i$, and let $b_{i,\mathcal{J}_i^c}\triangleq(b_{i,k}\in\{0,1\}:k\in\mathcal{J}_i^c)$ be a binary tuple. An $(n, \mathcal{J}_\mathcal{I}, b_{\mathcal{I},\mathcal{J}_\mathcal{I}^c})$-polar code, where $\mathcal{J}_\mathcal{I}\triangleq(\mathcal{J}_i : i\in\mathcal{I})$ and $b_{\mathcal{I},\mathcal{J}_\mathcal{I}^c}\triangleq(b_{i, \mathcal{J}_i^c}: i\in\mathcal{I})$,
 consists of the following:
% . An $(n, \mathcal{J})$-polar code with
%$
%\mathcal{J}^c\triangleq  \{1, 2, \ldots, n\}\setminus \mathcal{J}
%$
% consists of the following:
\begin{enumerate}
\item An index set for information bits transmitted by node~$i$ denoted by $\mathcal{J}_i$
%\begin{equation}
%\mathcal{J}_i\triangleq \left\{  k\in \{1, 2, \ldots, n\}\left|\parbox[c]{2.45 in}{$Z^{[p_{X_\mathcal{I}}; q_{Y|X_\mathcal{I}}]}(U_{i,k}|U_i^{k-1}, X_{[i-1]}^n, Y^n) \!\le \! \frac{1}{n^4}$}  \!\!\!\right.  \right\} \label{defInformationBitSetMAC}
%\end{equation}
for each $i\in\mathcal{I}$.
The set
$
 \mathcal{J}_i^c $
 is referred to as the index set for frozen bits transmitted by node~$i$.

 \item A message set $\mathcal{W}_i\triangleq \{1, 2, \ldots, 2^{|\mathcal{J}_i|}\}$ for each $i\in\mathcal{I}$, where~$W_i$ is uniform on~$\mathcal{W}_i$.

 \item An encoding bijection $f_i^{\text{MAC}}: \mathcal{W}_i \rightarrow \mathcal{U}_{i, \mathcal{J}_i}$ for encoding $W_i$ into $|\mathcal{J}_i|$ information bits denoted by $U_{i, \mathcal{J}_i}$ for each $i\in\mathcal{I}$ such that
   \begin{equation*}
  U_{i, \mathcal{J}_i} = f_i^{\text{MAC}}(W_i),
   \end{equation*}
   where $\mathcal{U}_{i, \mathcal{J}_i}$ and $  U_{i, \mathcal{J}_i}$ are defined as $\mathcal{U}_{i, \mathcal{J}_i} \triangleq \prod_{k\in\mathcal{J}_i}\mathcal{U}_k$ and $ U_{i, \mathcal{J}_i} \triangleq (U_{i,k}:k\in\mathcal{J}_i)$ respectively.
   Since message~$W_i$ is uniform on~$\mathcal{W}_i$, $f_i^{\text{MAC}}(W_i)$ is a sequence of i.i.d.\ uniform bits such that
   \begin{equation}
 \Pr\{U_{i, \mathcal{J}_i} = u_{i, \mathcal{J}_i}\}=\frac{1}{2^{|\mathcal{J}_i|}} \label{defUniformBitsMAC}
   \end{equation}
   for all $u_{i, \mathcal{J}_i}\in\{0,1\}^{|\mathcal{J}_i|}$, where the bits are transmitted through the polarized channels indexed by~$\mathcal{J}_i$.
For each $i\in\mathcal{I}$ and each $k\in \mathcal{J}_i^c$, let
\begin{equation}
U_{i,k}=b_{i,k} \label{frozenBitsMAC}
\end{equation}
be the frozen bit to be transmitted by node~$i$ in time slot~$k$.
After $U_i^n$ has been determined, node~$i$ transmits $X_i^n$ where
\begin{equation}
[X_{i,1}\ X_{i,2}\ \ldots \ X_{i,n}] \triangleq [U_{i,1}\ U_{i,2}\ \ldots \ U_{i,n}]G_n^{-1}.  \label{defpUgivenXinPolarMAC}
\end{equation}

\item A sequence of successive cancellation decoding functions $\varphi_{i,k}^{\text{MAC}}: \{0, 1\}^{k-1}\times \{0,1\}^{(i-1)n} \times \mathcal{Y}^n \rightarrow \{0,1\}$ for each $i\in\mathcal{I}$ and each $k\in\{1, 2, \ldots, n\}$ such that the recursively generated $(\hat U_{1,1}, \ldots, \hat U_{1,n}), (\hat U_{2,1}, \ldots, \hat U_{2,n}),\ldots, (\hat U_{N,1}, \ldots, \hat U_{N,n})$ and $(\hat X_{1,1}, \ldots, \hat X_{1,n}), (\hat X_{2,1}, \ldots, \hat X_{2,n}),\ldots, (\hat X_{N,1}, \ldots, \hat X_{N,n})$ are produced as follows. For each $i\in\mathcal{I}$ and each $k=1, 2, \ldots, n$, given that $\hat U_i^{k-1}$, $\hat U_{[i-1]}^n$ and $\hat X_{[i-1]}^n$ were constructed before the construction of $\hat U_{i,k}$, node~$\mathrm{d}$ constructs the estimate of $U_{i,k}$ through computing
    \begin{equation*}
    \hat U_{i,k} \triangleq \varphi_{i,k}^{\text{MAC}}(\hat U_i^{k-1}, \hat X_{[i-1]}^n, Y^n)
    \end{equation*}
    where
    \begin{align}
  \hat u_{i,k} &\triangleq  \varphi_{i,k}^{\text{MAC}}(\hat u_i^{k-1}, \hat x_{[i-1]}^n, y^n) \notag\\*
  &=
    \begin{cases}
  0  & \text{if $k\in\mathcal{J}_i$ and $p_{U_{i,k}|U_i^{k-1}, X_{[i-1]}^n, Y^n}(0|\hat u_i^{k-1}, \hat x_{[i-1]}^n, y^n) \ge p_{U_{i,k}|U_i^{k-1}, X_{[i-1]}^n, Y^n}(1|\hat u_i^{k-1}, \hat x_{[i-1]}^n, y^n)$,}\vspace{0.04 in}\\
   1 & \text{if $k\in\mathcal{J}_i$ and $p_{U_{i,k}|U_i^{k-1}, X_{[i-1]}^n, Y^n}(0|\hat u_i^{k-1}, \hat x_{[i-1]}^n, y^n)  < p_{U_{i,k}|U_i^{k-1}, X_{[i-1]}^n, Y^n}(1|\hat u_i^{k-1}, \hat x_{[i-1]}^n, y^n)$,}\vspace{0.04 in} \\
   b_{i,k} & \text{if $k\in\mathcal{J}_i^c$.}
    \end{cases}
    \label{defSCdecoderMAC}
    \end{align}
    After obtaining $\hat U_i^n$, node~$\mathrm{d}$ constructs the estimate of $X_i^n$ through computing
    \begin{align}
   [\hat X_{i,1}\ \hat X_{i,2}\ \ldots \ \hat X_{i,n}] \triangleq [\hat U_{i,1}\ \hat U_{i,2}\ \ldots \ \hat U_{i,n}]G_n^{-1}
    \end{align}
    and
    declares that
    \begin{equation*}
    \hat W_i \triangleq \left(f_i^{\text{MAC}}\right)^{-1}(\hat U_{i, \mathcal{J}_i})
    \end{equation*}
    is the transmitted~$W_i$ where $\left(f_i^{\text{MAC}}\right)^{-1}$ denote the inverse function of $f_i^{\text{MAC}}$.
\end{enumerate}
%If the sequence of encoding functions $f_k$ satisfies the EH constraints~\eqref{eqn:eh}, the code is also called an {\em $(n, \mathcal{J})$-EH polar code}.
\end{Definition}
\medskip
\begin{Remark}\label{remark4}
By inspecting Definition~\ref{defCodeMAC} and Definition~\ref{defPolarCodeWithFrozenBitsMAC}, we see that every $(n, \mathcal{J}_\mathcal{I}, b_{\mathcal{I}, \mathcal{J}_\mathcal{I}^c})$-polar code is also an \linebreak $\big(n, (2^{|\mathcal{J}_1|},2^{|\mathcal{J}_2|}, \ldots, 2^{|\mathcal{J}_N|})\big)$-code.
\end{Remark}
\medskip
\begin{Definition} \label{defPolarCodeMAC}
The uniform-input $(n, \mathcal{J}_\mathcal{I})$-polar code is defined as an $(n, \mathcal{J}_\mathcal{I}, B_{\mathcal{I},\mathcal{J}_\mathcal{I}^c})$-polar code where $B_{\mathcal{I},\mathcal{J}_\mathcal{I}^c}$ consists of i.i.d.\ uniform bits that are independent of the message~$W_\mathcal{I}$.
\end{Definition}
%\medskip
%\begin{Remark} \label{remark2}
%For the uniform-input $(n, \mathcal{J})$-polar code as defined in Definition~\ref{defPolarCode}, the Bhattacharyya parameters $Z^{[p_X^*; q_{Y|X}]}(U_k|U^{k-1}, Y^n)$ and $Z^{[p_X^*; q_{Y|X}]}(U_k|U^{k-1})$ are calculated according to $p_{U^n, X^n, Y^n}$ as stated in~\eqref{defP} where $p_{X^n}(x^n)=\prod_{k=1}^n p_X^*(x_k)$. On the other hand, it follows from Definition~\ref{defPolarCode} that $U^n$ generated from the $(n, \mathcal{J})$-polar code consists of~$n$ i.i.d.\ uniform bits, which together with the definition of the polarization map $G_n$ in~\eqref{defGn} implies that $p_{U^n, X^n, Y^n}$ is indeed the distribution induced by the polar code.
%\end{Remark}
%%\medskip
\medskip
\begin{Definition}\label{defErrorProbabilityPolarMAC}
For the uniform-input $(n, \mathcal{J}_\mathcal{I})$-polar code defined for the MAC, the probability of decoding error is defined as
\begin{equation*}
\Pr\{\hat W_\mathcal{I} \ne W_\mathcal{I}\} = \Pr\bigg\{\bigcup_{i\in\mathcal{I}} \{U_{i, \mathcal{J}_i} \ne \hat U_{i, \mathcal{J}_i}\}\bigg\}
\end{equation*}
where the error is averaged over the random messages and the frozen bits. The code is also called a uniform-input $(n, \mathcal{J}_\mathcal{I}, \varepsilon)$-polar code if the probability of decoding error is no larger than $\varepsilon$.
\end{Definition}
\medskip

%The following proposition is a generalization of Proposition~\ref{propositionErrorPolar}, and
The following proposition bounds the error probability in terms of Bhattacharyya parameters, and it is a generalization of the well-known result for the special case $N=1$~(e.g., see \cite[Proposition~2]{Arikan}). The proof of Proposition~\ref{propositionErrorPolarMAC} can be deduced from~\cite[Sec.~IV]{AbbeTelatar12}, and is contained in Appendix~\ref{appendixA} for completeness.
\medskip
\begin{Proposition} \label{propositionErrorPolarMAC}
For the uniform-input $(n, \mathcal{J}_\mathcal{I})$-polar code defined for the MAC $q_{Y|X_\mathcal{I}}$, we have
\begin{align}
\Pr\{\hat W_\mathcal{I} \ne W_\mathcal{I}\}  \le \sum_{i=1}^N\sum_{k\in \mathcal{J}_i} Z^{[p_{X_\mathcal{I}}; q_{Y|X_\mathcal{I}}]}(U_{i,k}|U_i^{k-1}, X_{[i-1]}^n, Y^n). \label{st1PropErrorPolarMAC}
\end{align}
\end{Proposition}
\medskip

The following proposition
%, which is a generalization of Proposition~\ref{propositionBMCscaling},
follows from combining Lemma~\ref{lemmaPolarMAC}, Definition~\ref{defSymmetricCapacityMAC} and Proposition~\ref{propositionErrorPolarMAC}.
\medskip
\begin{Proposition} \label{propositionPolarCodeMAC}
There exists a universal constant $t>0$ such that the following holds. Fix any $N$-source binary-input MAC characterized by $q_{Y|X_\mathcal{I}}$. Fix any $m\in\mathbb{N}$, let $n=2^m$ and define
\begin{equation}
\mathcal{J}_i^{\text{SE}}\triangleq \left\{k\in \{1, 2, \ldots, n\}\left|\parbox[c]{2.45 in}{$Z^{[p_{X_\mathcal{I}}^*; q_{Y|X_\mathcal{I}}]}(U_{i,k}|U_i^{k-1}, X_{[i-1]}^n, Y^n) \!\le \! \frac{1}{n^4}$}  \!\!\!\right.  \right\} \label{defInformationBitSetMAC}
%, \\ Z^{[p_X^*; q_{Y|X}]}(U_k|U^{k-1}) \ge 1-\frac{1}{n^4}
\end{equation}
for each $i\in\mathcal{I}$ where $p_{X_\mathcal{I}}^*$ is the uniform distribution as defined in~\eqref{defPI*} and the superscript ``SE" stands for ``scaling exponent". Then, the corresponding uniform-input $(n, \mathcal{J}_\mathcal{I}^{\text{SE}})$-polar code satisfies
\begin{align}
 \frac{\sum_{i=1}^N\left|\mathcal{J}_i^{\text{SE}}\right|}{n} \ge C_{\text{sum}} - \frac{tN}{n^{1/\beta}} \label{st1PropPolarCodeMAC}
\end{align}
and
\begin{align}
\Pr\{\hat W_\mathcal{I} \ne W_\mathcal{I}\}  \le \frac{N}{n^3}\,. \label{st2PropPolarCodeMAC}
\end{align}
\end{Proposition}
\begin{IEEEproof}
Let $t>0$ be the universal constant specified in Lemma~\ref{lemmaPolarMAC} and fix an~$n$.
%\begin{align}
%\mathcal{J}_i\triangleq\left\{ k\in\{1, 2, \ldots, n\}\left|\parbox[c]{2.45 in}{$ Z^{[p_{X_\mathcal{I}}; q_{Y|X_\mathcal{I}}]}(U_{i,k}|U_i^{k-1}, X_{[i-1]}^n, Y^n)  \le \frac{1}{n^4}$}  \right.  \right\}
% \end{align}
%for each $i\in\mathcal{I}$, and let
 For each $i\in\mathcal{I}$, it follows from Lemma~\ref{lemmaPolarMAC} and Proposition~\ref{propositionErrorPolarMAC} that
\begin{align}
 \frac{\left|\mathcal{J}_i^{\text{SE}}\right|}{n} \ge I_{p_{X_\mathcal{I}}^* q_{Y|X_\mathcal{I}}}(X_i; X_{[i-1]},Y) - \frac{t}{n^{1/\beta}} \label{eq1PropPolarCodeMAC}
\end{align}
and
\begin{align}
\Pr\{\hat W_\mathcal{I} \ne W_\mathcal{I}\}  \le \frac{N}{n^3} \label{eq2PropPolarCodeMAC}
\end{align}
for the uniform-input $(n, \mathcal{J}_\mathcal{I}^{\text{SE}})$-polar code. Since $p_{X_\mathcal{I}}^*=\prod_{i=1}^N p_{X_i}^*$, it follows that
\begin{equation}
I_{p_{X_\mathcal{I}}^* q_{Y|X_\mathcal{I}}}(X_i; X_{[i-1]},Y) = I_{p_{X_\mathcal{I}}^* q_{Y|X_\mathcal{I}}}(X_i; Y|X_{[i-1]})
\end{equation}
holds for each $i\in\mathcal{I}$, which implies that
\begin{equation}
\sum_{i=1}^N I_{p_{X_\mathcal{I}}^* q_{Y|X_\mathcal{I}}}(X_i; X_{[i-1]},Y) = I_{p_{X_\mathcal{I}}^* q_{Y|X_\mathcal{I}}}(X_\mathcal{I}; Y). \label{eq3PropPolarCodeMAC}
\end{equation}
Consequently, \eqref{st1PropPolarCodeMAC} follows from~\eqref{eq1PropPolarCodeMAC}, \eqref{eq3PropPolarCodeMAC} and Definition~\ref{defSymmetricCapacityMAC}, and \eqref{st2PropPolarCodeMAC} follows from~\eqref{eq2PropPolarCodeMAC}.
\end{IEEEproof}
\medskip
\begin{Remark}
Proposition~\ref{propositionPolarCodeMAC} shows that the sum-capacity of a binary-input MAC with $N$ sources can be achieved within a gap of $O(N n^{-1/\beta})$ by using a superposition of $N$ binary-input polar codes.
\end{Remark}

\section{Problem Formulation of the AWGN Channel and New Polarization Results}
\label{sectionDefinitionAWGN}
\subsection{The AWGN Channel} \label{subsec2G}
%In order to achieve the capacity of the additive white Gaussian noise (AWGN) channel, we would like to adopt
%The main result of this paper is to show that uniform-input polar codes for binary-input MACs described in Definition~\ref{defPolarCodeWithFrozenBitsMAC}. To this end,
It is well known that appropriately designed polar codes are capacity-achieving for the AWGN channel~\cite{AbbeBarron11}. The main contribution of this paper is proving an upper bound on the scaling exponent of polar codes for the AWGN channel by using uniform-input polar codes for binary-input MACs described in Definition~\ref{defPolarCodeWithFrozenBitsMAC}.
%The channel model of the AWGN channel is described as follows.
%The AWGN channel is defined to be a point-to-point memoryless channel consisting of a source $\mathrm{s}$ and a destination~$\mathrm{d}$ as described in Section~\ref{subsecPTP} with the following additional assumptions:
%\begin{enumerate}
%\item $\mathcal{X}=\mathcal{Y}=\mathbb{R}$.
%\item The channel law is $Y=X+Z$ where $Z$ is the standard normal random variable that is independent of~$X$.
%\item For any $n\in\mathbb{N}$, the following peak power constraint has to be satisfied for some admissible power~$P>0$:
%\begin{equation}
%\Pr\left\{\frac{1}{n}\sum_{k=1}^n X_k^2 \le P\right\}=1. \label{peakPowerConstraint}
%\end{equation}
%\end{enumerate}
The following two definitions formally define the AWGN channel and length-$n$ codes for the channel.
\medskip
\begin{Definition}\label{defCodeAWGN}
An {\em $(n, M, P)$-code} is an $(n, M)$-code described in Definition~\ref{defCode} subject to the additional assumptions that
$\mathcal{X}=\mathbb{R}$ and the peak power constraint
\begin{equation}
\Pr\left\{\frac{1}{n}\sum_{k=1}^n X_k^2 \le P\right\}=1 \label{peakPowerConstraint}
\end{equation}
 is satisfied.
\end{Definition}
\medskip
\begin{Definition}\label{defChannelAWGN}
The {\em AWGN channel} is a point-to-point memoryless channel described in Definition~\ref{defChannel} subject to the additional assumption that $\mathcal{Y}=\mathbb{R}$ and
$
q_{Y|X}(y|x)=\mathcal{N}(y; x, 1)$
for all $x\in \mathbb{R}$ and $y\in \mathbb{R}$.
%Since $p_{Y_k|X_k}$ does not depend on~$i$ by \eqref{defChannelInDefinition*}, the channel is stationary.
\end{Definition}
\medskip
\begin{Definition} \label{defErrorProbabilityAWGN}
For an $(n, M, P)$-code defined on the AWGN channel, we can calculate according to~\eqref{memorylessStatement} the \textit{average probability of error} defined as $\Pr\big\{\hat W \ne W\big\}$.
We call an $(n, M, P)$-code with average probability of error no larger than~$\varepsilon$ an {\em $(n, M, P, \varepsilon)$-code}.
\end{Definition}
%\medskip
%\begin{Definition} \label{defAchievableRateAWGN}
%Let $\varepsilon\in (0,1)$ be a real number. A rate $R$ is said to be \textit{$\varepsilon$-achievable} for the AWGN channel if there exists a sequence of $(n, M_n, P, \varepsilon)$-codes such that
%\begin{equation*}
%\liminf_{n\rightarrow \infty}\frac{1}{n}\log M_n \ge R.
%\end{equation*}
%\end{Definition}
%\medskip
%\begin{Definition}\label{defCapacityAWGN}
%Let $\varepsilon\in (0,1)$ be a real number. The {\em $\varepsilon$-capacity} of the AWGN channel, denoted by $C_\varepsilon$, is defined to be
%\[
%C_\varepsilon^{\text{AWGN}} \triangleq \sup\{R\in\mathbb{R}_+\,| R\text{ is $\varepsilon$-achievable for the AWGN channel}\}.
% \]
% The \emph{capacity} of the AWGN channel is $C^{\text{AWGN}}\triangleq \inf_{\varepsilon>0}C_\varepsilon^{\text{AWGN}}$. Recall the definition of~$\mathrm{C}(P)$ in~\eqref{defCP}. It is well known that the AWGN channel possesses the strong converse property~\cite{Sha59b}, i.e.,
%\begin{equation}
%C_\varepsilon^{\text{AWGN}}=\mathrm{C}(P) %\label{intro:strongConverse}
%\end{equation}
%for each $\varepsilon\in(0,1)$.
%\end{Definition}
 \subsection{Uniform-Input Polar Codes for the AWGN Channel} \label{subsec2H}
Recall that we would like to use uniform-input polar codes for binary-input MACs described in Definition~\ref{defPolarCodeWithFrozenBitsMAC} to achieve the capacity of the AWGN channel, i.e., $\mathrm{C}(P)$ in~\eqref{defCP}. The following definition describes the basic structure of such uniform-input polar codes.
\medskip
\begin{Definition} \label{defPolarCodeAWGNavg}
Fix an $n=2^m$ where $m\in \mathbb{N}$.
%For each $i\in\{1, 2, \ldots, m\}$, let $\mathcal{J}_i\subseteq \{1, 2, \ldots, n\}$ be a set to be specified shortly and define $\mathcal{J}_i^c \triangleq \{1, 2, \ldots, n\}\setminus\mathcal{J}_i$.
%, and let $b_{i,\mathcal{J}_i^c}\triangleq(b_{i,k}\in\{0,1\}:k\in\mathcal{J}_i^c)$ be a binary tuple.
An $(n, \mathcal{J}_\mathcal{I}, P, \mathcal{A})_{\text{avg}}$-polar code with average power~$P$ and input alphabet $\mathcal{A}$ consists of the following:
% . An $(n, \mathcal{J})$-polar code with
%$
%\mathcal{J}^c\triangleq  \{1, 2, \ldots, n\}\setminus \mathcal{J}
%$
% consists of the following:
\begin{enumerate}
\item An input alphabet $\mathcal{A}\subset\mathbb{R}\cup\{0^-\}$ with $|\mathcal{A}|=n$ such that
\begin{equation}
 \frac{1}{n}\sum_{a\in\mathcal{A}\setminus \{0^-\}}a^2 \le P,
 \end{equation}
 where $\mathbb{R}\cup\{0^-\}$ can be viewed as a line with $2$ origins.\footnote{Introducing the symbol $0^-$ allows us to create a set of cardinality~$n$ which consists of $n-2$ non-zero real numbers and $2$ origins $0$ and $0^-$}
We index each element of $\mathcal{A}$ by a unique length-$m$ binary tuple
 \begin{equation}
 x_\mathcal{I}^{\text{MAC}}\triangleq(x_1^{\text{MAC}}, x_2^{\text{MAC}}, \ldots, x_m^{\text{MAC}}),
 \end{equation}
 and let $\rho: \{0,1\}^m\rightarrow \mathcal{A}$ be the bijection that maps the indices to the elements of $\mathcal{A}$ such that
$
  \rho(x_\mathcal{I}^{\text{MAC}})$ denotes
  the element in $\mathcal{A}$ indexed by~$x_\mathcal{I}^{\text{MAC}}$.

 \item A binary-input MAC $q_{Y|X_\mathcal{I}^{\text{MAC}}}$ induced by $\mathcal{A}$ as defined through Definitions~\ref{defChannelMAC} and~\ref{defChannelBinaryMAC} with the identifications $N=m$ and $\mathcal{I}=\{1, 2, \ldots, m\}$.

 \item A message set $\mathcal{W}_i\triangleq \{1, 2, \ldots, 2^{|\mathcal{J}_i|}\}$ for each $i\in\{1, 2, \ldots, m\}$, where~$\mathcal{W}_\mathcal{I}$ is the message alphabet of the uniform-input $(n, \mathcal{J}_\mathcal{I})$-polar code for the binary-input MAC $q_{Y|X_\mathcal{I}^{\text{MAC}}}$ as defined through Definitions~\ref{defPolarCodeWithFrozenBitsMAC} and~\ref{defPolarCodeMAC} such that
     \begin{equation}
     |\mathcal{W}_\mathcal{I}|= \prod_{i=1}^m |\mathcal{W}_i|=2^{\sum_{i=1}^m |\mathcal{J}_i|}.
     \end{equation}
In addition, $W_\mathcal{I}$ is uniform on~$\mathcal{W}_\mathcal{I}$. We view the uniform-input $(n, \mathcal{J}_\mathcal{I})$-polar code as an $\big(n, (2^{|\mathcal{J}_1|},2^{|\mathcal{J}_2|}, \ldots, 2^{|\mathcal{J}_N|})\big)$-code (cf.\ Remark~\ref{remark4}) and let $\{f_{i,k}^{\text{MAC}}\,|\,i\in\mathcal{I}, k\in\{1, 2, \ldots, n\}\}$ and $\varphi^{\text{MAC}}$ denote the corresponding set of encoding functions and the decoding function respectively (cf.\ Definition~\ref{defCodeMAC}).

\item An encoding function $f_k: \mathcal{W}_\mathcal{I} \rightarrow \mathcal{A}$
defined as
\begin{equation}
 f_k(W_\mathcal{I}) \triangleq \rho(f_{1,k}^{\text{MAC}}(W_1), f_{2,k}^{\text{MAC}}(W_2),\ldots,f_{m,k}^{\text{MAC}}(W_m))
   \end{equation}
for each $k\in\{1, 2, \ldots, n\}$, where $f_k$ is used for encoding $W_\mathcal{I}$ into $X_k$ such that
   \begin{equation}
  X_k =
  \begin{cases}
   f_k(W_\mathcal{I}) &\text{if $ f_k(W_\mathcal{I})\ne 0^-$,}\\
   0 & \text{if $ f_k(W_\mathcal{I})=0^-$.}
   \end{cases} \label{defEncodingAWGN}
   \end{equation}
   Note that both the encoded symbols $0$ and $0^-$ in $\mathcal{A}$ result in the same transmitted symbol $0\in\mathbb{R}$ according to~\eqref{defEncodingAWGN}.
By construction, $ f_1(W_\mathcal{I})$, $ f_2(W_\mathcal{I})$, $\ldots$, $ f_n(W_\mathcal{I})$ are i.i.d.\ random variables that are uniformly distributed on $\mathcal{A}$ and hence $X_1$, $X_2$, $\ldots$, $X_n$ are i.i.d.\ real-valued random variables (but not necessarily uniform).
 %\item An encoding function $f_i^{\text{MAC}}: \mathcal{W}_i \rightarrow \mathcal{X}_i^n$ for encoding $W_i$ into $X_i^n$ for each $i\in\mathcal{I}$ such that
%   \begin{equation*}
%  X_i^n = f_i^{\text{MAC}}(W_i).
%   \end{equation*}

\item A decoding function $\varphi: \mathbb{R}^n \rightarrow \mathcal{W}_\mathcal{I}$ defined as
\begin{equation}
\varphi \triangleq \varphi^{\text{MAC}}
\end{equation}
such that
\begin{equation}
\hat{\mathcal{W}}_\mathcal{I} = \varphi(Y^n).
\end{equation}
\end{enumerate}
%If the sequence of encoding functions $f_k$ satisfies the EH constraints~\eqref{eqn:eh}, the code is also called an {\em $(n, \mathcal{J})$-EH polar code}.
\end{Definition}
\medskip
\begin{Remark}
For an $(n, \mathcal{J}_\mathcal{I}, P, \mathcal{A})_{\text{avg}}$-polar code, the flexibility of allowing $\mathcal{A}$ to contain~$2$ origins is crucial to proving the main result of this paper. This is because the input distribution which we will use to establish scaling results for the AWGN channel in Theorem~\ref{thmMainResult} can be viewed as the uniform distribution over some set that contains~$2$ origins, although the input distribution in the real domain as specified in~\eqref{defDistPprime} to follow is not uniform.
 %^More specifically, we will construct a good $\mathcal{A}$ that contains~$2$ origins in the following proposition and lemma followed by using the good $\mathcal{A}$ to establish scaling results for the AWGN channel .
\end{Remark}
\medskip
\begin{Proposition}\label{propositionPolarAWGNavg}
There exists a universal constant $t>0$ such that the following holds. Suppose we are given an $(n, \mathcal{J}_\mathcal{I}^{\text{SE}}, P, \mathcal{A})_{\text{avg}}$-polar code defined for the AWGN channel $q_{Y|X}$ with a $2$-origin $\mathcal{A}$ (i.e., $\mathcal{A}\supseteq\{0, 0^-\})$. Define $\mathcal{X}\triangleq \mathcal{A}\setminus \{0^-\} \subset \mathbb{R}$ where $\mathcal{X}$ contains $1$ origin and $n-2$ non-zero real numbers. Then, the $(n, \mathcal{J}_\mathcal{I}^{\text{SE}}, P, \mathcal{A})_{\text{avg}}$-polar code is an $(n, M)$-code (cf.\ Definition~\ref{defCode}) which satisfies
\begin{align}
 \frac{1}{n}\log M \ge I_{p_{X}^\prime q_{Y|X}}(X;Y) - \frac{t \log n}{n^{1/\beta}}, \label{st1propositionPolarAWGNavg}
\end{align}
\begin{align}
\Pr\{\hat W_\mathcal{I} \ne W_\mathcal{I}\}  \le \frac{\log n}{n^3}\,, \label{st2propositionPolarAWGNavg}
\end{align}
and
\begin{align}
\Pr\left\{X^n = x^n\right\}= \prod_{k=1}^n \Pr\left\{X_k = x_k\right\}= \prod_{k=1}^n p_X^\prime(x_k)
\end{align}
%\begin{align}
%\E\left[\frac{1}{n}\sum_{k=1}^n X_k^2\right]\le P
%\end{align}
for all $x^n\in\mathcal{X}^n$
%and
%\begin{equation}
%\E_{p_X^\prime}[X^2]\le P
%\end{equation}
where $p_X^\prime$ is the distribution on $\mathcal{X}$ defined as
\begin{equation}
p_{X}^\prime(a)=\begin{cases}\frac{1}{n} &\text{if $a\ne 0$,} \\
\frac{2}{n} & \text{if $a=0$.}\end{cases} \label{defDistPprime}
\end{equation}
\end{Proposition}
\begin{IEEEproof}
The proposition follows from inspecting Proposition~\ref{propositionPolarCodeMAC} and Definition~\ref{defPolarCodeAWGNavg} with the identifications $N=\log n$ and $\log M = \sum_{i=1}^m |\mathcal{J}_i|$.
\end{IEEEproof}
\medskip
%\begin{Definition}\label{defQuantileDistribution}
%Let
%\begin{equation}
%\mathcal{D}_n \triangleq \left\{\left.\frac{k}{n} + \frac{1}{2n}\,\right|\,k\in\{0, 1, \ldots, n-1\}\right\}.
%\end{equation}
%%and let $U_n$ be uniformly distributed on~$\mathcal{D}_n$.
%Define the set
%\begin{equation}
%\mathcal{A}_n = \sqrt{P-\frac{1}{n^{1/3}}}\,\Phi^{-1}(\mathcal{D}_n).
%\end{equation}
%\end{Definition}
%\medskip

The following lemma, a strengthened version of~\cite[Th.~6]{AbbeBarron11}, provides a construction of a good $\mathcal{A}$ which leads to a controllable gap between $\mathrm{C}(P)$ and $I_{p_X^\prime q_{Y|X}}(X;Y)$ for the corresponding $(n, \mathcal{J}_\mathcal{I}^{\text{SE}}, P, \mathcal{A})_{\text{avg}}$-polar code. Although the following lemma is intuitive, the proof is technical and hence relegated to Appendix~\ref{appendixB}.
\medskip
\begin{Lemma} \label{lemmaGoodA}
Let $q_{Y|X}$ be the conditional distribution that characterizes the AWGN channel, and fix any $\gamma\in[0,1)$. For each $n=2^m$ where $m\in \mathbb{N}$,
define
\begin{equation}
\mathcal{D}_n \triangleq \left\{\left.\frac{\ell}{n} \,\right|\,\ell\in\bigg\{1, 2, \ldots, \frac{n}{2}, \ldots n-1\bigg\}\right\}, \label{defSetDn}
\end{equation}
define
\begin{equation}
s_X(x)\triangleq \mathcal{N}\left(x; 0, P\Big(1-\frac{1}{n^{(1-\gamma)/\beta}}\Big)\right) \label{defDistS}
\end{equation}
 for all $x\in\mathbb{R}$, define $\Phi_X$ to be the cdf of $s_X$,
%and let $U_n$ be uniformly distributed on~$\mathcal{D}_n$.
and define
\begin{equation}
\mathcal{X} \triangleq \Phi_X^{-1}(\mathcal{D}_n). \label{defSetX}
\end{equation}
%there exists a set
%\begin{equation}
%\mathcal{A}\triangleq \{a_1, a_2, \ldots, a_n\} \label{defSetA}
%\end{equation}
%consisting of~$n$ real numbers such that the following holds:
Note that $\mathcal{X}$ contains $1$ origin and $n-2$ non-zero real numbers, and we let $p_X^\prime$ be the distribution on~$\mathcal{X}$ as defined in~\eqref{defDistPprime}. In addition, define the distribution $p_{X^n}^\prime(x^n)\triangleq \prod_{k=1}^n p_X^\prime(x_k)$. Then, there exists a constant $t^\prime >0$ that depends on~$P$ and~$\gamma$ but not~$n$ such that the following statements hold for each $n\in\mathbb{N}$:
\begin{align}
\E_{p_X^\prime}[X^2]\le P\Big(1-\frac{1}{n^{(1-\gamma)/\beta}}\Big), \label{lemmaGoodAst1}
\end{align}
%\begin{align}
%\Var_{p_X^\prime}[X^2]\le 3P^2, \label{lemmaGoodAst2}
%\end{align}
%\begin{align}
%\E_{p_X^\prime}\left[e^{\frac{X^2}{2\sqrt{n} P\big(1-\frac{1}{n^{(1-\gamma)/\beta}}\big)}}\right] =\left(1-\frac{1}{\sqrt{n}}\right)^{-1/2}\, , \label{lemmaGoodAst2}
%\end{align}
\begin{align}
\Pr_{p_{X^n}^\prime}\left\{\frac{1}{n}\sum_{k=1}^n X_k^2 >P\right\} \le \frac{e^3}{e^{n^{\frac{1}{2}-\frac{1-\gamma}{\beta}}}}\, , \label{lemmaGoodAst2}
\end{align}
and
\begin{align}
\mathrm{C}(P)- I_{p_X^\prime q_{Y|X}}(X;Y) \le \frac{t^\prime \sqrt{\log n}}{n^{(1-\gamma)/\beta}}. \label{lemmaGoodAst3}
\end{align}
\end{Lemma}
\medskip

%The following theorem is a direct consequence of Proposition~\ref{propositionPolarAWGNavg} and Lemma~\ref{lemmaGoodA}.
%\medskip
%\begin{Theorem} \label{thmPolarAWGNavg}
%There exists a constant $t^*>0$ that depends on~$P$ but not~$n$ such that the following holds. For any $n\in\mathbb{N}$, there exists an $\mathcal{A}$ such that the corresponding $(n, \mathcal{J}_\mathcal{I}^{\text{SE}}, P, \mathcal{A})_{\text{avg}}$-polar code defined for the AWGN channel $q_{Y|X}$ satisfies
%\begin{align}
% \frac{1}{n}\sum_{i=1}^N |\mathcal{J}_i^{\text{SE}}| \ge \mathrm{C}(P)-\frac{t^* \log n}{n^{1/\beta}}, \label{st1thmPolarAWGNavg}
%\end{align}
%\begin{align}
%\Pr\{\hat W_\mathcal{I} \ne W_\mathcal{I}\}  \le \frac{\log n}{n^3}\,, \label{st2thmPolarAWGNavg}
%\end{align}
%\begin{align}
%\E\left[ \frac{1}{n}\sum_{k=1}^n X_k^2\right]\le P\Big(1-\frac{1}{n^{1/\beta}}\Big)  \label{st3thmPolarAWGNavg}
%\end{align}
%and
%\begin{align}
%\Pr\left\{\frac{1}{n}\sum_{k=1}^n X_k^2 >P\right\} \le \frac{e^3}{e^{n^{\frac{1}{2}-\frac{1}{\beta}}}}\,.  \label{st4thmPolarAWGNavg}
%\end{align}
%%\begin{align}
%%\E_{p_X^\prime}\left[e^{\frac{X^2}{2n P\big(1-\frac{1}{n^{1/3}}\big)}}\right] = \sqrt{1-\frac{1}{n}}\, , \label{lemmaGoodAst2}
%%\end{align}
%%\begin{align}
%%\E\left[\frac{1}{n}\sum_{k=1}^n X_k^2\right]\le P
%%\end{align}
%\end{Theorem}
%\medskip

A shortcoming of Proposition~\ref{propositionPolarAWGNavg} is that the $(n, \mathcal{J}_\mathcal{I}^{\text{SE}}, P, \mathcal{A})_{\text{avg}}$-polar code may not satisfy the peak power constraint~\eqref{peakPowerConstraint} and hence it may not qualify as an $(n, 2^{\sum_{i=1}^N |\mathcal{J}_i^{\text{SE}}|}, P)$-code (cf.\ Definition~\ref{defCodeAWGN}). Therefore, we describe in the following definition a slight modification of an $(n, \mathcal{J}_\mathcal{I}^{\text{SE}}, P, \mathcal{A})_{\text{avg}}$-polar code so that the modified polar code always satisfies the peak power constraint~\eqref{peakPowerConstraint}.
\medskip
\begin{Definition} \label{defPolarCodeAWGNpeak}
The \emph{$0$-power-outage version} of an $(n, \mathcal{J}_\mathcal{I}, P, \mathcal{A})_{\text{avg}}$-polar code is an $(n, 2^{\sum_{i=1}^N |\mathcal{J}_i|})$-code which follows identical encoding and decoding operations of the $(n, \mathcal{J}_\mathcal{I}, P, \mathcal{A})_{\text{avg}}$-polar code except that the source will modify the input symbol in a time slot~$k$ if the following scenario occurs: Let $X^n$ be the desired codeword generated by the source according to the encoding operation of the $(n, \mathcal{J}_\mathcal{I}, P, \mathcal{A})_{\text{avg}}$-polar code, where the randomness of $X^n$ originates from the information bits $(U_{i, \mathcal{J}_i}: i\in\mathcal{I})$ and the frozen bits $B_{\mathcal{I}, \mathcal{J}_\mathcal{I}^c}$. If transmitting the desired symbol~$X_k$ at time~$k$ results in violating the power constraint~$\frac{1}{n}\sum_{\ell=1}^k  X_\ell^2 > P$, the source will transmit the symbol~$0$ at time~$k$ instead. An $(n, 2^{\sum_{i=1}^N |\mathcal{J}_i|}, \varepsilon)$-code is called an $(n, \mathcal{J}_\mathcal{I}, P, \mathcal{A}, \varepsilon)_{\text{peak}}$-polar code if it is the $0$-power-outage version of some $(n, \mathcal{J}_\mathcal{I}, P, \mathcal{A})_{\text{avg}}$-polar code.
\end{Definition}
\medskip

By Definition~\ref{defPolarCodeAWGNpeak}, every $(n, \mathcal{J}_\mathcal{I}, P, \mathcal{A}, \varepsilon)_{\text{peak}}$-polar code satisfies the peak power constraint~\eqref{peakPowerConstraint} and hence achieves zero power outage, i.e., $\Pr\{\frac{1}{n}\sum_{k=1}^n X_k^2 > P\}=0$. Using Definition~\ref{defPolarCodeAWGNpeak}, we obtain the following corollary which states the obvious fact that the probability of power outage of an $(n, \mathcal{J}_\mathcal{I}, P, \mathcal{A})_{\text{avg}}$-polar code can be viewed as part of the probability of error of the $0$-power-outage version of the code.
\medskip
\begin{Corollary} \label{corollaryMainResultAWGN}
Given an $(n, \mathcal{J}_\mathcal{I}, P, \mathcal{A})_{\text{avg}}$-polar code, define
\begin{equation}
\varepsilon_1\triangleq\Pr\{\hat W_\mathcal{I} \ne W_\mathcal{I}\}
\end{equation}
and
\begin{equation}
\varepsilon_2\triangleq\Pr\left\{\frac{1}{n}\sum_{k=1}^n X_k^2> P\right\}.
\end{equation}
Then, the $0$-power-outage version of the $(n, \mathcal{J}_\mathcal{I}, P, \mathcal{A})_{\text{avg}}$-polar code is an $(n, \mathcal{J}_\mathcal{I}, P, \mathcal{A}, \varepsilon_1+\varepsilon_2)_{\text{peak}}$-polar code that satisfies the peak power constraint~\eqref{peakPowerConstraint}.
\end{Corollary}

%\medskip
%
%The following theorem is a direct consequence of Theorem~\ref{thmPolarAWGNavg} and Corollary~\ref{corollaryMainResultAWGN}.
%\medskip
%\begin{Theorem}\label{thmMainResult*}
%There exists a constant $t^*>0$ that depends on~$P$ but not~$n$ such that the following holds. For any $n\in\mathbb{N}$, there exists an $\mathcal{A}$ such that the corresponding $(n, \mathcal{J}_\mathcal{I}^{\text{SE}}, P, \mathcal{A})_{\text{peak}}$-polar code defined for the AWGN channel $q_{Y|X}$ satisfies
%\eqref{st1thmPolarAWGNavg} and
%\begin{align}
%\Pr\{\hat W_\mathcal{I} \ne W_\mathcal{I}\}  \le \frac{\log n}{n^3} + \frac{e^3}{e^{n^{\frac{1}{2}-\frac{1}{\beta}}}}\,. \label{st2thmPolarAWGNavg}
%\end{align}
%\end{Theorem}
%\medskip
%\begin{Remark}
%A candidate of $\mathcal{A}$ in Theorem~\ref{thmMainResult*} can been explicitly constructed according to Lemma~\ref{lemmaGoodA} by the identification $\gamma\equiv 0$ and $\mathcal{A}\equiv \mathcal{X}\cup \{0^-\}$.
%\end{Remark}
\section{Scaling Exponents and Main Result} \label{sectionMainResult}

\subsection{Scaling Exponent of Uniform-Input Polar Codes for MACs}
%\begin{Definition} \label{defScalingExp}
%Fix an~$\varepsilon\in(0,1)$ and a BMC $q_{Y|X}$ with symmetric capacity $C_\text{sym}$ (cf.\ Definition~\ref{defSymmetricCapacity}).
%%\begin{equation}
%%I(q_{Y|X_\mathcal{I}})\triangleq \max\limits_{p_{X_\mathcal{I}}: \, p_{X_\mathcal{I}}= \prod\limits_{i\in\mathcal{I}}p_{X_i}}I_{p_{X_\mathcal{I}}q_{Y|X_\mathcal{I}}}(X_\mathcal{I}; Y).
%%\end{equation}
%% Define
%%\begin{equation}
%%M_{\text{\tiny PC-BMSC}}^*(n, \varepsilon) \triangleq \max\{M_n|\,\text{There exists an $(n, M_n, \varepsilon)$-polar code on $q_{Y|X}$}\}
%%\end{equation}
%% for each $n\in\mathbb{N}$.
%The \emph{scaling exponent of uniform-input polar codes for the BMC} is defined as
%\begin{align*}
%\mu_\varepsilon^{\text{\tiny PC-BMC}} \triangleq \liminf_{m\rightarrow \infty} \inf_{\mathcal{J}} \left\{\left.\frac{-\log n}{\log \left|C_\text{sym}- \frac{|\mathcal{J}|}{n}\right|}\right| \text{ $n=2^m$, there exists a uniform-input $(n, \mathcal{J}, \varepsilon)$-polar code on $q_{Y|X}$}\right\}.
%\end{align*}
%\end{Definition}
% , and equal to~$0$ for $\varepsilon\in [1/2,1)$ for all DMCs (the optimal non-asymptotic achievable rates exceed capacity for non-degenerate DMCs for $\varepsilon>1/2$)}.
% For a general BMC which does not need to be symmetric, we will see later in Lemma~\ref{lemmaBMCupperBound}, a stepping stone for establishing our main result, that the upper bound 4.714 in~$\eqref{muKnown}$ continues to hold.
We define scaling exponent of uniform-input polar codes for the binary-input MAC as follows.
 \medskip
\begin{Definition} \label{defScalingExpMAC}
Fix an~$\varepsilon\in(0,1)$ and an $N$-source binary-input MAC $q_{Y|X_\mathcal{I}}$ with symmetric sum-capacity $C_\text{sum}$ (cf.\ Definition~\ref{defSymmetricCapacityMAC}).
%\begin{equation}
%I(q_{Y|X_\mathcal{I}})\triangleq \max\limits_{p_{X_\mathcal{I}}: \, p_{X_\mathcal{I}}= \prod\limits_{i\in\mathcal{I}}p_{X_i}}I_{p_{X_\mathcal{I}}q_{Y|X_\mathcal{I}}}(X_\mathcal{I}; Y).
%\end{equation}
% Define
%\begin{equation}
%M_{\text{\tiny PC-BMSC}}^*(n, \varepsilon) \triangleq \max\{M_n|\,\text{There exists an $(n, M_n, \varepsilon)$-polar code on $q_{Y|X}$}\}
%\end{equation}
% for each $n\in\mathbb{N}$.
The \emph{scaling exponent of uniform-input polar codes for the MAC} is defined as
\begin{align*}
\mu_\varepsilon^{\text{\tiny PC-MAC}} \triangleq \liminf_{m\rightarrow \infty} \inf_{\mathcal{J}_\mathcal{I}} \left\{\left.\frac{-\log n}{\log \left|C_\text{sum}- \frac{\sum_{i=1}^N|\mathcal{J}_i|}{n}\right|}\right| \text{$n=2^m$, there exists a uniform-input $(n, \mathcal{J}_\mathcal{I}, \varepsilon)$-polar code on $q_{Y|X_\mathcal{I}}$}\right\}.
\end{align*}
\end{Definition}
%
%Definition~\ref{defScalingExp} formalizes the notion that we are seeking the smallest $\beta\ge 0$ such that $|I(q_{Y|X})-R_n|=O(n^{-1/\beta})$ holds.
%It has been shown in \cite[Sec.~IV-C]{HAU14} and \cite[Th.~2]{MHU15} that
%\begin{equation}
%3.579 \le \mu_\varepsilon^{\text{\tiny PC-BMC}}\le 4.714  \qquad \forall \varepsilon\in(0,1)\label{muKnown}
% \end{equation}
% for any BMSC $q_{Y|X}$ with capacity $I(q_{Y|X})\in (0,1)$. We note from~\cite[Th.~48]{PPV10} (also \cite{Strassen} and~\cite{Hayashi09}) that the \emph{optimal} scaling exponents (optimized over all codes) are equal to~$2$ for $\varepsilon\in (0, 1/2)$ for non-degenerate DMCs.
%% , and equal to~$0$ for $\varepsilon\in [1/2,1)$ for all DMCs (the optimal non-asymptotic achievable rates exceed capacity for non-degenerate DMCs for $\varepsilon>1/2$)}.
% For a general BMC which does not need to be symmetric, we will see later in Lemma~\ref{lemmaBMCupperBound}, a stepping stone for establishing our main result, that the upper bound 4.714 in~$\eqref{muKnown}$ continues to hold. In this paper, we are interested in the scaling exponent of save-and-transmit polar codes for the binary-input EH channel, which is formally defined as follows.
\medskip

Definition~\ref{defScalingExpMAC} formalizes the notion that we are seeking the smallest $\mu\ge 0$ such that $|C_\text{sum}-R_n^{\text{MAC}}|=O\big(n^{-\frac{1}{\mu}}\big)$ holds where $R_n^{\text{MAC}}\triangleq \frac{|\mathcal{J}|}{n}$ denotes the rate of an $(n, \mathcal{J}_\mathcal{I}, \varepsilon)$-polar code.
It has been shown in \cite[Sec.~IV-C]{HAU14} and \cite[Th.~2]{MHU15} that
\begin{equation}
3.579 \le \mu_\varepsilon^{\text{\tiny PC-BMC}}\le \beta = 4.714 \qquad \forall \varepsilon\in(0,1)\label{muKnown}
 \end{equation}
% for any BMC whose symmetric capacity lies in $(0,1)$.
for the special case $N=1$ where the binary-input MAC reduces a BMC.
 We note from~\cite[Th.~48]{PPV10} (also \cite{Strassen} and~\cite{Hayashi09}) that the \emph{optimal} scaling exponent (optimized over all codes) for any non-degenerate DMC (as well as BMC) is equal to~$2$ for all $\varepsilon\in (0, 1/2)$.

Using Proposition~\ref{propositionErrorPolarMAC} and Definition~\ref{defScalingExpMAC}, we obtain the following corollary, which shows that 4.714, the upper bound on $\mu_\varepsilon^{\text{\tiny PC-BMC}}$ in~\eqref{muKnown} for BMCs, remains to be a valid upper bound on the scaling exponent for binary-input MACs.
\medskip
\begin{Corollary} \label{corollayMainResultMAC}
Fix any $\varepsilon\in(0,1)$ and any binary-input MAC $q_{Y|X_\mathcal{I}}$. Then,
\begin{equation*}
\mu_\varepsilon^{\text{\tiny PC-MAC}}\le \beta = 4.714. \label{st1ThmMainResultMAC}
\end{equation*}
%More specifically, there exists a universal constant $t>0$ that does not depend on~$\varepsilon$ and~$N$ such that the following holds for any sufficiently large~$k\in\mathbb{N}$: Let $n=2^k$. Then, we can construct a uniform-input $(n, \mathcal{J}_\mathcal{I}, \varepsilon)$-polar code on $q_{Y|X_\mathcal{I}}$ such that
%\begin{equation}
%\frac{\sum_{i=1}^N|\mathcal{J}_i|}{n} \ge C_\text{sum}-  \frac{t N}{n^{1/\beta}}. \label{st2ThmMainResultMAC}
%\end{equation}
\end{Corollary}
%\begin{IEEEproof}
%Using Proposition~\ref{propositionPolarCodeMAC}, we have~\eqref{st2ThmMainResultMAC}. In addition, \eqref{st1ThmMainResultMAC} follows directly from~\eqref{st2ThmMainResultMAC} and Definition~\ref{defScalingExpMAC}.
%\end{IEEEproof}

%EH channel in spite of the additional EH constraints~\eqref{EHconstraintPolar}. The proof of the main result will be provided in Section~\ref{sectionProofofMainResult}.
%\medskip
%\begin{Theorem} \label{thmMainResult}
%For any $\varepsilon\in(0,1)$ and any binary-input EH channel,
%\begin{equation*}
%\mu_\varepsilon^{\text{\tiny PC-EH}}\le 4.714.
%\end{equation*}
%\end{Theorem}
%
%Theorem~\ref{thmMainResult} states that $4.714$ remains to be a valid upper bound on the scaling exponent of polar codes for the binary-input EH channel. This implies that the EH constraints do not worsen the rate of convergence to capacity if polar codes are employed. The chief intuition of this result is the following: We design the length of the saving phase~$m$ sufficiently small so that the convergence rate to the capacity $I(q_{Y|X})$ is not affected. Yet, this choice of~$m$ ensures that the probability that the EH constraints are violated is small (cf.\ Proposition~\ref{propositionCharacteristicFunctionDMC}), and essentially does not significantly worsen the overall probability of decoding error. An auxiliary contribution of this paper is that the upper bound on the scaling exponent holds for binary-input memoryless asymmetric channels, which is established in Lemma~\ref{lemmaBMCupperBound} as an important step to proving Theorem~\ref{thmMainResult}.

\subsection{Scaling Exponent of Uniform-Input Polar Codes for the AWGN channel}

\begin{Definition} \label{defScalingExpAWGN}
Fix a $P>0$ and an~$\varepsilon\in(0,1)$.
The \emph{scaling exponent of uniform-input polar codes for the AWGN channel} is defined as
\begin{align*}
\mu_{P,\varepsilon}^{\text{\tiny PC-AWGN}} \triangleq \liminf_{m\rightarrow \infty} \inf_{\mathcal{J}_\mathcal{I}, \mathcal{A}} \left\{\left.\frac{-\log n}{\log \left|\mathrm{C}(P)- \frac{\sum_{i=1}^N|\mathcal{J}_i|}{n}\right|}\right| \text{$n=2^m$, there exists a uniform-input $(n, \mathcal{J}_\mathcal{I}, P, \mathcal{A}, \varepsilon)_{\text{peak}}$-polar code}\right\}.
\end{align*}
\end{Definition}
\medskip

Definition~\ref{defScalingExpAWGN} formalizes the notion that we are seeking the smallest $\mu\ge 0$ such that $\left|\mathrm{C}(P)-R_n^{\text{AWGN}}\right|=O\big(n^{-\frac{1}{\mu}}\big)$ holds where $R_n^{\text{AWGN}}\triangleq \frac{\sum_{i=1}^N|\mathcal{J}_i|}{n}$ denotes the rate of an $(n, \mathcal{J}_\mathcal{I}, P, \varepsilon)_{\text{peak}}$-polar code.
 We note from~\cite[Th.~54]{PPV10} and~\cite[Th.~5]{Hayashi09}
%in view of~\eqref{eqn:asymp_expans}
that the \emph{optimal} scaling exponent of the optimal code for the AWGN channel is equal to~$2$ for any $\varepsilon\in (0, 1/2)$.
%We note from~\cite[Th.~48]{PPV10} (also \cite{Strassen} and~\cite{Hayashi09})
% , and equal to~$0$ for $\varepsilon\in [1/2,1)$ for all DMCs (the optimal non-asymptotic achievable rates exceed capacity for non-degenerate DMCs for $\varepsilon>1/2$)}. %
%It has been shown in \cite[Sec.~IV-C]{HAU14} and \cite[Th.~2]{MHU15} that
%\begin{equation*}
%3.579 \le \mu_\varepsilon^{\text{\tiny PC-BMC}}\le 4.714%\label{muKnown}
% \end{equation*}
% for any BMC $q_{Y|X}$ with capacity $I(q_{Y|X})\in (0,1)$ and any $\varepsilon\in(0,1)$.
The following theorem is the main result of this paper, which shows that $4.714$ is a valid upper bound on the scaling exponent of polar codes for the AWGN channel.
\medskip
\begin{Theorem} \label{thmMainResult}
Fix any $P>0$ and any $\varepsilon\in(0,1)$. There exists a constant $t^*>0$ that does not depend on~$n$ such that the following holds. For any $n=2^m$ where $m\in \mathbb{N}$, there exists an $\mathcal{A}$ such that the corresponding $(n, \mathcal{J}_\mathcal{I}^{\text{SE}}, P, \mathcal{A})_{\text{peak}}$-polar code defined for the AWGN channel $q_{Y|X}$ satisfies
\begin{align}
 \frac{1}{n}\sum_{i=1}^N |\mathcal{J}_i^{\text{SE}}| \ge \mathrm{C}(P)-\frac{t^* \log n}{n^{1/\beta}}, \label{st1thmPolarAWGNavg}
\end{align}
 and
\begin{align}
\Pr\{\hat W_\mathcal{I} \ne W_\mathcal{I}\}  \le \frac{\log n}{n^3} + \frac{e^3}{e^{n^{\frac{1}{2}-\frac{1}{\beta}}}}\,. \label{st2thmPolarAWGNavg}
\end{align}
%For any $P>0$ and any $\varepsilon\in(0,1)$,
In particular, we have
\begin{equation}
\mu_{P,\varepsilon}^{\text{\tiny PC-AWGN}}\le \beta =  4.714. \label{stThmMainResult}
\end{equation}
\end{Theorem}
\begin{IEEEproof}
%Fix any $P>0$ and any $\varepsilon\in(0,1)$. %In addition, fix any $m\in\mathbb{N}$ and let $n=2^m$.
%Proposition~\ref{propositionPolarAWGNavg} and Lemma~\ref{lemmaGoodA}.
%\medskip
%\begin{Theorem} \label{thmPolarAWGNavg}
%There exists a constant $t^*>0$ that depends on~$P$ but not~$n$ such that the following holds. For any $n\in\mathbb{N}$, there exists an $\mathcal{A}$ such that the corresponding $(n, \mathcal{J}_\mathcal{I}^{\text{SE}}, P, \mathcal{A})_{\text{avg}}$-polar code defined for the AWGN channel $q_{Y|X}$ satisfies
%\begin{align}
% \frac{1}{n}\sum_{i=1}^N |\mathcal{J}_i^{\text{SE}}| \ge \mathrm{C}(P)-\frac{t^* \log n}{n^{1/\beta}}, \label{st1thmPolarAWGNavg}
%\end{align}
%\begin{align}
%\Pr\{\hat W_\mathcal{I} \ne W_\mathcal{I}\}  \le \frac{\log n}{n^3}\,, \label{st2thmPolarAWGNavg}
%\end{align}
%\begin{align}
%\E\left[ \frac{1}{n}\sum_{k=1}^n X_k^2\right]\le P\Big(1-\frac{1}{n^{1/\beta}}\Big)  \label{st3thmPolarAWGNavg}
%\end{align}
%and
%\begin{align}
%\Pr\left\{\frac{1}{n}\sum_{k=1}^n X_k^2 >P\right\} \le \frac{e^3}{e^{n^{\frac{1}{2}-\frac{1}{\beta}}}}\,.  \label{st4thmPolarAWGNavg}
%\end{align}
%Let $t>0$ be the constant that appears in Proposition~\ref{propositionPolarAWGNavg}.
Fix a $P>0$, an $\varepsilon\in(0,1)$ and an $n=2^m$ where $m\in \mathbb{N}$. Combining Proposition~\ref{propositionPolarAWGNavg} and Lemma~\ref{lemmaGoodA}, we conclude that there exist a constant $t^*>0$ that does not depend on~$n$ and an $\mathcal{A}$ such that the corresponding $(n, \mathcal{J}_\mathcal{I}^{\text{SE}}, P, \mathcal{A})_{\text{avg}}$-polar code defined for the AWGN channel $q_{Y|X}$ satisfies~\eqref{st1thmPolarAWGNavg}
%\begin{align}
%\Pr\{\hat W_\mathcal{I} \ne W_\mathcal{I}\}  \le \frac{\log n}{n^3}\,, \label{st2thmPolarAWGNavgProof}
%\end{align}
%%\begin{align}
%%\E\left[ \frac{1}{n}\sum_{k=1}^n X_k^2\right]\le P\Big(1-\frac{1}{n^{1/\beta}}\Big)  \label{st3thmPolarAWGNavgProof}
%%\end{align}
%%and
%and
%\begin{align}
%\Pr\left\{\frac{1}{n}\sum_{k=1}^n X_k^2 >P\right\} \le \frac{e^3}{e^{n^{\frac{1}{2}-\frac{1}{\beta}}}}\,.  \label{st4thmPolarAWGNavg}
%\end{align}
%It follows from Theorem~\ref{thmMainResult*} that for any $n=2^m$ where $m\in\mathbb{N}$, there exist an $\mathcal{A}$ and the corresponding
%$(n, \mathcal{J}_\mathcal{I}^{\text{SE}}, P, \mathcal{A})_{\text{avg}}$-polar code that satisfies
%\begin{align}
% \frac{1}{n}\sum_{i=1}^N |\mathcal{J}_i^{\text{SE}}| \ge \mathrm{C}(P)-\frac{t^* \log n}{n^{1/\beta}}, \label{st1thmPolarAWGNavgMainResult}
%\end{align}
\begin{align}
\Pr\{\hat W_\mathcal{I} \ne W_\mathcal{I}\}  \le \frac{\log n}{n^3}\,, \label{st2thmPolarAWGNavgMainResult}
\end{align}
%\begin{align}
%\E\left[ \frac{1}{n}\sum_{k=1}^n X_k^2\right]\le P\Big(1-\frac{1}{n^{1/\beta}}\Big)  \label{st3thmPolarAWGNavgMainResult}
%\end{align}
and
\begin{align}
\Pr\left\{\frac{1}{n}\sum_{k=1}^n X_k^2 >P\right\} \le \frac{e^3}{e^{n^{\frac{1}{2}-\frac{1}{\beta}}}}\,.  \label{st4thmPolarAWGNavgMainResult}
\end{align}
%\eqref{st1thmPolarAWGNavg}, \eqref{st2thmPolarAWGNavg}, \eqref{st3thmPolarAWGNavg} and \eqref{st4thmPolarAWGNavg}.
%Using \eqref{st3thmPolarAWGNavg}, \eqref{st4thmPolarAWGNavg} and the Markov's inequality, we have
%\begin{align}
%&\Pr\left\{\frac{1}{n}\sum_{k=1}^n X_k^2 > P\right\} \notag\\
%&\quad = \Pr\left\{\frac{\sum_{k=1}^n X_k^2}{2\sqrt{n}P\big(1-\frac{1}{n^{1/3}}\big)} > \frac{\sqrt{n}}{2\big(1-\frac{1}{n^{1/3}}\big)}\right\} \\
%&\quad = \Pr\left\{\frac{\sum_{k=1}^n X_k^2}{2\sqrt{n}P\big(1-\frac{1}{n^{1/3}}\big)} > \frac{\sqrt{n}}{2\big(1-\frac{1}{n^{1/3}}\big)}\right\}\\
%&\quad \le \frac{}{} \\
%&\quad \le \frac{3P^2}{n^{1/3}}. \label{st5thmPolarAWGNavg}
%\end{align}
Using \eqref{st2thmPolarAWGNavgMainResult}, \eqref{st4thmPolarAWGNavgMainResult} and Corollary~\ref{corollaryMainResultAWGN}, we conclude that the $(n, \mathcal{J}_\mathcal{I}^{\text{SE}}, P, \mathcal{A})_{\text{avg}}$-polar code is an
$(n, \mathcal{J}_\mathcal{I}^{\text{SE}}, P, \mathcal{A})_{\text{peak}}$-polar code that satisfies~\eqref{st1thmPolarAWGNavg} and~\eqref{st2thmPolarAWGNavg}. Since
\begin{align}
 \frac{\log n}{n^3}+\frac{e^3}{e^{n^{\frac{1}{2}-\frac{1}{\beta}}}} \le \varepsilon  \label{st6thmPolarAWGNavg}
\end{align}
for all sufficiently large~$n$, it follows from~\eqref{st1thmPolarAWGNavg}, \eqref{st2thmPolarAWGNavg} and Definition~\ref{defScalingExpAWGN} that~\eqref{stThmMainResult} holds.
%Consequently, \eqref{stThmMainResult} follows from~\eqref{}
\end{IEEEproof}
\section{Moderate Deviations Regime} \label{sectionModerateDeviation}
\subsection{Polar Codes That Achieve the Symmetric Capacity of a BMC} \label{MD:subsec2C}
%\subsection{Polarization for Binary-Input Memoryless Channels} \label{MD:subsec2B}

%  It is well known that \cite[Proposition~2]{Arikan:10ISIT}
 % where
%  \begin{equation}
%  p_{Y^n|X^n}(y^n|x^n)=\prod_{k=1}^n q_{Y|X}(y_k|x_k) \label{MD:defPYnGivenXn}
%  \end{equation}
The following result is based on~\cite[Sec.~IV]{MHU15},  which developed a tradeoff between the gap to capacity and the decay rate of the error probability for a BMC under the moderate deviations regime~\cite{altug14b} where both the gap to capacity and the error probability vanish as~$n$ grows.
%The proof combines key ideas in~\cite{MHU15} and~\cite{HondaYamamoto13}, and is relegated to Appendix~\ref{MD:appendixA}.
\medskip
\begin{Lemma}[{\cite[Sec.~IV]{MHU15}}]\label{MD:lemmaPolar}
%Let $\beta=4.714$ and
There exists a universal constant $t_{\text{MD}}>0$ such that the following holds. Fix any $\gamma\in\left(\frac{1}{1+\beta},1\right)$ and any BMC characterized by $q_{Y|X}$. Recall that $p_X^*$ denotes the uniform distribution on $\{0,1\}$. Then for any $n=2^m$ where $m\in \mathbb{N}$, we have
 \begin{align}
 \frac{1}{n}\left|\left\{ k\in\{1, 2, \ldots, n\}\left|\parbox[c]{2.7 in}{$ Z^{[p_X^*; q_{Y|X}]}(U_k|U^{k-1}, Y^n) \le 2^{-n^{\gamma h_2^{-1}\left(\frac{\gamma\beta + \gamma -1}{\gamma \beta}\right)}}$} \!\! \right.  \right\}\right| \ge I_{p_X^* q_{Y|X}}(X; Y) - \frac{t_{\text{MD}}}{n^{(1-\gamma)/\beta}} \label{MD:st1InLemmaPolar}
 \end{align}
where $h_2:[0, 1/2]\rightarrow[0,1]$ denotes the binary entropy function.
%and
%  \begin{align*}
%& \frac{1}{n}\!\left|\left\{ k\in\{1, 2, \ldots, n\}\!\left| \parbox[c]{2.18 in}{$ Z_{p_{U^n, X^n, Y^n}}(U_k|U^{k-1}, Y^n) \ge 1-  \frac{1}{n^4}, \\Z_{p_{U^n, X^n, Y^n}}(U_k|U^{k-1}) \le \frac{1}{n^4}$} \!\!\right.  \right\}\right| \notag\\
% &\quad\ge 1- I_{p_X q_{Y|X}}(X; Y) - \frac{t_2}{n^{1/\beta}}. % \label{MD:st2InLemmaPolar}
% \end{align*}
\end{Lemma}

\subsection{Polar Codes that Achieve the Symmetric Sum-Capacity of a Binary-Input MAC} \label{MD:subsec2F}
The following lemma, whose proof is omitted because it is analogous to the proof of Lemma~\ref{lemmaPolarMAC}, is a direct consequence of Lemma~\ref{MD:lemmaPolar}.
%%The proof combines key ideas in~\cite{MHU15} and~\cite{HondaYamamoto13}, and is relegated to Appendix~\ref{MD:appendixA}.
\medskip
\begin{Lemma}\label{MD:lemmaPolarMAC}
There exists a universal constant $t_{\text{MD}}>0$ such that the following holds. Fix any $\gamma\in\left(\frac{1}{1+\beta},1\right)$ and any binary-input MAC characterized by $q_{Y|X_\mathcal{I}}$. Recall that $p_{X_\mathcal{I}}^*=\prod_{i\in\mathcal{I}}p_{X_i}^*$. Then for any $n=2^m$ where $m\in \mathbb{N}$, we have
 \begin{align}
 &\frac{1}{n}\left|\left\{ k\in\{1, 2, \ldots, n\}\left|\text{$ Z^{[p_{X_\mathcal{I}}^*; q_{Y|X_\mathcal{I}}]}(U_{i,k}|U_i^{k-1}, X_{[i-1]}^n, Y^n)  \le 2^{-n^{\gamma h_2^{-1}\left(\frac{\gamma\beta + \gamma -1}{\gamma \beta}\right)}}$}  \right.  \right\}\right| \notag\\
&\quad\ge I_{p_{X_\mathcal{I}}^* q_{Y|X_\mathcal{I}}}(X_i; X_{[i-1]},Y) - \frac{t_{\text{MD}}}{n^{(1-\gamma)/\beta}} \label{MD:st1InLemmaPolarMAC}
 \end{align}
 for each $i\in\mathcal{I}$.
% \begin{align}
% \frac{1}{n}\left|\left\{ k\in\{1, 2, \ldots, n\}\left|\parbox[c]{1.9 in}{$ Z^{[p_X; q_{Y|X}]}(U_k|U^{k-1}, Y^n) \le \frac{1}{n^4}, \vspace{0.04 in} \\
% Z^{[p_X; q_{Y|X}]}(U_k|U^{k-1}) \ge 1-\frac{1}{n^4}$}  \right.  \right\}\right| \ge I_{p_{X_\mathcal{I}} q_{Y|X_\mathcal{I}}}(X_i; X_{[i-1]},Y) - \frac{t}{n^{1/\beta}}. \label{MD:st1InLemmaPolar}
% \end{align}
%and
%  \begin{align*}
%& \frac{1}{n}\!\left|\left\{ k\in\{1, 2, \ldots, n\}\!\left| \parbox[c]{2.18 in}{$ Z_{p_{U^n, X^n, Y^n}}(U_k|U^{k-1}, Y^n) \ge 1-  \frac{1}{n^4}, \\Z_{p_{U^n, X^n, Y^n}}(U_k|U^{k-1}) \le \frac{1}{n^4}$} \!\!\right.  \right\}\right| \notag\\
% &\quad\ge 1- I_{p_X q_{Y|X}}(X; Y) - \frac{t_2}{n^{1/\beta}}. % \label{MD:st2InLemmaPolar}
% \end{align*}
\end{Lemma}
%\begin{IEEEproof}
%Fix any $i\in\mathcal{I}$. Construct $p_{X_{[i-1]}, Y|X_i}$ by marginalizing $p_{X_\mathcal{I}} q_{Y|X_\mathcal{I}}$ and view $p_{X_{[i-1]}, Y|X_i}$ as the conditional distribution that characterizes a BMC. The lemma then follows directly from Lemma~\ref{MD:lemmaPolar}.
%\end{IEEEproof}
\medskip

Combining Lemma~\ref{MD:lemmaPolarMAC}, Definition~\ref{defSymmetricCapacityMAC} and Proposition~\ref{propositionErrorPolarMAC},  we obtain the following  proposition, whose proof is analogous to the proof of Proposition~\ref{propositionPolarCodeMAC} and hence omitted.
\medskip
\begin{Proposition} \label{MD:propositionPolarCodeMAC}
There exists a universal constant $t_{\text{MD}}>0$ such that the following holds.  Fix any $\gamma\in\left(\frac{1}{1+\beta},1\right)$ and any $N$-source binary-input MAC characterized by $q_{Y|X_\mathcal{I}}$. In addition, fix any $m\in\mathbb{N}$, let $n=2^m$ and define
\begin{equation}
\mathcal{J}_i^{\text{MD}}\triangleq \left\{k\in \{1, 2, \ldots, n\}\left|\text{$Z^{[p_{X_\mathcal{I}}; q_{Y|X_\mathcal{I}}]}(U_{i,k}|U_i^{k-1}, X_{[i-1]}^n, Y^n) \!\le \! 2^{-n^{\gamma h_2^{-1}\left(\frac{\gamma\beta + \gamma -1}{\gamma \beta}\right)}}$}  \right.  \right\} \label{MD:defInformationBitSetMAC}
%, \\ Z^{[p_X^*; q_{Y|X}]}(U_k|U^{k-1}) \ge 1-\frac{1}{n^4}
\end{equation}
for each $i\in\mathcal{I}$ where the superscript ``MD" stands for ``moderate deviations". Then, the corresponding uniform-input $(n, \mathcal{J}_\mathcal{I}^{\text{MD}})$-polar code described in Definition~\ref{defPolarCodeMAC} satisfies
\begin{align}
 \frac{\sum_{i=1}^N\left|\mathcal{J}_i^{\text{MD}}\right|}{n} \ge C_{\text{sum}} - \frac{t_{\text{MD}}N}{n^{(1-\gamma)/\beta}} \label{MD:st1PropPolarCodeMAC}
\end{align}
and
\begin{align}
\Pr\{\hat W_\mathcal{I} \ne W_\mathcal{I}\}  \le  Nn\, 2^{-n^{\gamma h_2^{-1}\left(\frac{\gamma\beta + \gamma -1}{\gamma \beta}\right)}}. \label{MD:st2PropPolarCodeMAC}
\end{align}
\end{Proposition}

 \subsection{Uniform-Input Polar Codes for the AWGN Channel} \label{MD:subsec2H}
\begin{Proposition}\label{MD:propositionPolarAWGNavg}
There exists a universal constant $t_{\text{MD}}>0$ such that the following holds. Fix any $\gamma\in\left(\frac{1}{1+\beta},1\right)$. Suppose we are given an $(n, \mathcal{J}_\mathcal{I}^{\text{MD}}, P, \mathcal{A})_{\text{avg}}$-polar code (cf.\ Definition~\ref{defPolarCodeAWGNavg}) defined for the AWGN channel $q_{Y|X}$ with a $2$-origin $\mathcal{A}$ (i.e., $\mathcal{A}\supseteq\{0, 0^-\})$. Define $\mathcal{X}\triangleq \mathcal{A}\setminus \{0^-\} \subset \mathbb{R}$ where $\mathcal{X}$ contains $1$ origin and $n-2$ non-zero real numbers. Then, the $(n, \mathcal{J}_\mathcal{I}^{\text{MD}}, P, \mathcal{A})_{\text{avg}}$-polar code is an $(n, M)$-code (cf.\ Definition~\ref{defCode}) which satisfies
\begin{align}
 \frac{1}{n}\log M \ge I_{p_{X}^\prime q_{Y|X}}(X;Y) - \frac{t_{\text{MD}} \log n}{n^{(1-\gamma)/\beta}}, \label{MD:st1propositionPolarAWGNavg}
\end{align}
\begin{align}
\Pr\{\hat W_\mathcal{I} \ne W_\mathcal{I}\}  \le  (\log n)n \,2^{-n^{\gamma h_2^{-1}\left(\frac{\gamma\beta + \gamma -1}{\gamma \beta}\right)}}, \label{MD:st2propositionPolarAWGNavg}
\end{align}
and
\begin{align}
\Pr\left\{X^n = x^n\right\}= \prod_{k=1}^n \Pr\left\{X_k = x_k\right\}= \prod_{k=1}^n p_X^\prime(x_k)
\end{align}
%\begin{align}
%\E\left[\frac{1}{n}\sum_{k=1}^n X_k^2\right]\le P
%\end{align}
for all $x^n\in\mathcal{X}^n$
%and
%\begin{equation}
%\E_{p_X^\prime}[X^2]\le P
%\end{equation}
where $p_X^\prime$ is the distribution on $\mathcal{X}$ as defined in \eqref{defDistPprime}.
\end{Proposition}
\begin{IEEEproof}
The proposition follows from inspecting Proposition~\ref{MD:propositionPolarCodeMAC} and Definition~\ref{defPolarCodeAWGNavg} with the identifications $N=\log n$ and $\log M = \sum_{i=1}^m |\mathcal{J}_i^{\text{MD}}|$.
\end{IEEEproof}
\medskip

The following theorem develops the tradeoff between the gap to capacity and the decay rate of the error probability for  $(n, \mathcal{J}_\mathcal{I}^{\text{MD}}, P, \mathcal{A})_{\text{peak}}$-polar codes defined for the AWGN channel.
\medskip
\begin{Theorem}\label{MD:mainResult}
Fix a $\gamma\in\left(\frac{1}{1+\beta},1\right)$. There exists a constant $t_{\text{MD}}^*>0$ that depends on~$P$ and~$\gamma$ but not~$n$ such that the following holds for any $n=2^m$ where $m\in\mathbb{N}$. There exists an $(n, \mathcal{J}_\mathcal{I}^{\text{MD}}, P, \mathcal{A}, \varepsilon)_{\text{peak}}$-polar code defined for the AWGN channel $q_{Y|X}$ that satisfies
\begin{align}
 \frac{1}{n}\sum_{i=1}^N |\mathcal{J}_i^{\text{MD}}| \ge \mathrm{C}(P)-\frac{t_{\text{MD}}^* \log n}{n^{(1-\gamma)/\beta}}, \label{MD:st1thmPolarAWGNavg}
\end{align}
and
\begin{align}
\Pr\{\hat W_\mathcal{I} \ne W_\mathcal{I}\}  \le  (n\log n + e^3) 2^{-n^{\gamma h_2^{-1}\left(\frac{\gamma\beta + \gamma -1}{\gamma \beta}\right)}}.
%+  \frac{e^3}{e^{n^{\frac{1}{2}-\frac{1-\gamma}{\beta}}}}.
\label{MD:st2thmPolarAWGNavg*}
\end{align}
%
%\begin{align}
%\Pr\left\{\frac{1}{n}\sum_{k=1}^n X_k^2 >P\right\} \le \,.  \label{MD:st4thmPolarAWGNavg*}
%\end{align}
\end{Theorem}
\begin{IEEEproof}
%\medskip
%\begin{Theorem} \label{MD:thmPolarAWGNavg}
%Fix a $\gamma\in\left(\frac{1}{1+\beta},1\right)$. There exists a constant $t_{\text{MD}}^*>0$ that depends on~$P$ and~$\gamma$ but not~$n$ such that the following holds for any $n=2^m$ where $m\in\mathbb{N}$. There exists an $\mathcal{A}$ such that the corresponding $(n, \mathcal{J}_\mathcal{I}^{\text{MD}}, P, \mathcal{A})_{\text{avg}}$-polar code defined for the AWGN channel $q_{Y|X}$ satisfies
%%\begin{align}
%% \frac{1}{n}\sum_{i=1}^N |\mathcal{J}_i^{\text{MD}}| \ge \mathrm{C}(P)-\frac{t_{\text{MD}}^* \log n}{n^{(1-\gamma)/\beta}}, \label{MD:st1thmPolarAWGNavg}
%%\end{align}
%%\begin{align}
%%\Pr\{\hat W_\mathcal{I} \ne W_\mathcal{I}\}  \le  (\log n)n 2^{-n^{\gamma h_2^{-1}\left(\frac{\gamma\beta + \gamma -1}{\gamma \beta}\right)}}, \label{MD:st2thmPolarAWGNavg}
%%\end{align}
%%\begin{align}
%%\E\left[ \frac{1}{n}\sum_{k=1}^n X_k^2\right]\le P\Big(1-\frac{1}{n^{(1-\gamma)/\beta}}\Big)  \label{MD:st3thmPolarAWGNavg}
%%\end{align}
%%and
%%\begin{align}
%%\Pr\left\{\frac{1}{n}\sum_{k=1}^n X_k^2 >P\right\} \le \frac{e^3}{e^{n^{\frac{1}{2}-\frac{1-\gamma}{\beta}}}}\,.  \label{MD:st4thmPolarAWGNavg}
%%\end{align}
%%\begin{align}
%%\E_{p_X^\prime}\left[e^{\frac{X^2}{2n P\big(1-\frac{1}{n^{1/3}}\big)}}\right] = \sqrt{1-\frac{1}{n}}\, , \label{lemmaGoodAst2}
%%\end{align}
%%\begin{align}
%%\E\left[\frac{1}{n}\sum_{k=1}^n X_k^2\right]\le P
%%\end{align}
%\end{Theorem}
%\medskip
By Proposition~\ref{MD:propositionPolarAWGNavg} and Lemma~\ref{lemmaGoodA},
%By Theorem~\ref{MD:thmPolarAWGNavg},
there exists a constant $t_{\text{MD}}^*>0$ that depends on~$P$ and~$\gamma$ but not~$n$ such that for any $n=2^m$ where $m\in\mathbb{N}$, there exist an $\mathcal{A}$ and the corresponding $(n, \mathcal{J}_\mathcal{I}^{\text{MD}}, P, \mathcal{A})_{\text{avg}}$-polar code that satisfies~\eqref{MD:st1thmPolarAWGNavg},
\begin{align}
\Pr\{\hat W_\mathcal{I} \ne W_\mathcal{I}\}  \le  (\log n)n \,2^{-n^{\gamma h_2^{-1}\left(\frac{\gamma\beta + \gamma -1}{\gamma \beta}\right)}}, \label{MD:st2thmPolarAWGNavg}
\end{align}
%\begin{align}
%\E\left[ \frac{1}{n}\sum_{k=1}^n X_k^2\right]\le P\Big(1-\frac{1}{n^{(1-\gamma)/\beta}}\Big)  \label{MD:st3thmPolarAWGNavg}
%\end{align}
and
\begin{align}
\Pr\left\{\frac{1}{n}\sum_{k=1}^n X_k^2 >P\right\} \le \frac{e^3}{e^{n^{\frac{1}{2}-\frac{1-\gamma}{\beta}}}}\,.  \label{MD:st4thmPolarAWGNavg}
\end{align}
It remains to show~\eqref{MD:st2thmPolarAWGNavg*}.
Using \eqref{MD:st2thmPolarAWGNavg}, \eqref{MD:st4thmPolarAWGNavg} and Corollary~\ref{corollaryMainResultAWGN}, we conclude that the $(n, \mathcal{J}_\mathcal{I}^{\text{MD}}, P, \mathcal{A})_{\text{avg}}$-polar code is an $(n, \mathcal{J}_\mathcal{I}^{\text{MD}}, P, \mathcal{A}, \varepsilon)_{\text{peak}}$-polar code that satisfies
\begin{align}
\varepsilon &\triangleq (n\log n ) 2^{-n^{\gamma h_2^{-1}\left(\frac{\gamma\beta + \gamma -1}{\gamma \beta}\right)}}+\frac{e^3}{e^{n^{\frac{1}{2}-\frac{1-\gamma}{\beta}}}} \\
& \le  (n\log n + e^3) 2^{-n^{\gamma h_2^{-1}\left(\frac{\gamma\beta + \gamma -1}{\gamma \beta}\right)}} %\label{MD:st6thmPolarAWGNavg}
\end{align}
where the inequality follows from the fact that $h_2(x) \ge 2x$ for all $x\in[0, 1/2]$. This concludes the proof.
%which together with~\eqref{st1thmPolarAWGNavgMainResult} implies that~\eqref{MD:st2thmPolarAWGNavg} holds.
\end{IEEEproof}
\medskip
\begin{Remark}
A candidate of $\mathcal{A}$ in Theorem~\ref{MD:mainResult} can been explicitly constructed according to Lemma~\ref{lemmaGoodA} by the identification $\mathcal{A}\equiv \mathcal{X}\cup \{0^-\}$.
\end{Remark}
%\medskip
%
%The authors in~\cite[Sec.~IV]{MHU15} provided an explicit construction of polar codes for any BMC which obey a certain tradeoff between the gap to capacity and the decay rate of error probability. More specifically, if the gap to capacity is set to vanish at a rate as slow as $\Theta\left(n^{-\frac{1-\gamma}{\beta}}\right)$ for some $\gamma\in\left(\frac{1}{1+\beta}, 1\right)$, then a length-$n$ polar code can be constructed such that the error probability is~$O\left(n \cdot 2^{-n^{\gamma h_2^{-1}\left(\frac{\gamma\beta + \gamma -1}{\gamma \beta}\right)}}\right)$ where $h_2:[0, 1/2]\rightarrow[0,1]$ denotes the binary entropy function. This tradeoff was developed under the moderate deviations regime~\cite{altug14b} where both the gap to capacity and the error probability vanish as~$n$ grows. For the AWGN channel, we develop a similar tradeoff under the moderate deviations regime by using our constructed polar codes described above.

\section{Concluding Remarks} \label{sectionConclusion}
In this paper, we provided an upper bound on the scaling exponent of polar codes for the AWGN channel (Theorem~\ref{thmMainResult}). In addition, we have shown in Theorem~\ref{MD:mainResult} a moderate deviations result --- namely, the existence of polar codes which obey a certain tradeoff between the gap to capacity and the decay rate of the error probability for the AWGN channel.

Since the encoding and decoding complexities of the binary-input polar code for a BMC are $O(n \log n)$ as long as we allow pseudorandom numbers to be shared between the encoder and the decoder for encoding and decoding the randomized frozen bits (e.g., see~\cite[Sec.~IX]{Arikan}), the encoding and decoding complexities of the polar codes for the AWGN channel defined in Definition~\ref{defPolarCodeAWGNavg} and Definition~\ref{defPolarCodeAWGNpeak} are $O(n \log n) \times \log n = O\big(n \log^2 n \big)$. By a standard probabilistic argument, there must exist a deterministic encoder for the frozen bits such that the decoding error of the polar code for the AWGN channel with the deterministic encoder is no worse than the polar code with randomized frozen bits. In the future, it may be fruitful to develop low-complexity algorithms for finding a good deterministic encoder for encoding the frozen bits. Another interesting direction for future research is to compare the empirical performance between our polar codes in Definitions~\ref{defPolarCodeAWGNavg} and~\ref{defPolarCodeAWGNpeak} and the state-of-the-art polar codes. One may also explore various techniques (e.g., list decoding, cyclic redundancy check (CRC), etc.) to improve the empirical performance of the polar codes constructed herein.

\appendices
\section{Proof of Proposition~\ref{propositionErrorPolarMAC}} \label{appendixA}%
Unless specified otherwise, all the probabilities in this proof are evaluated according to the distribution induced by the uniform-input $(n, \mathcal{J}_\mathcal{I})$-polar code. Consider
\begin{align}
\Pr\{\hat W_\mathcal{I} \ne W_\mathcal{I}\}  &= \sum_{i=1}^N \Pr\big\{\{\hat W_i \ne W_i\} \cap \{\hat W_{[i-1]} = W_{[i-1]}\}\big\} \\*
&  = \sum_{i=1}^N \Pr\big\{\{\hat U_i^n \ne U_i^n\} \cap \{\hat X_{[i-1]}^n = X_{[i-1]}^n\}\big\} \label{eq1aPropositionErrorPolarMAC}\\
& = \sum_{i=1}^N \sum_{k=1}^n \Pr\big\{\{\hat U_{i,k} \ne U_{i,k}\}\cap \{\hat U_i^{k-1}=U_i^{k-1}\} \cap \{\hat X_{[i-1]}^n = X_{[i-1]}^n\}\big\} \label{eq1PropositionErrorPolarMAC}
\end{align}
where
\eqref{eq1aPropositionErrorPolarMAC} is due to the fact by Definition~\ref{defPolarCodeWithFrozenBitsMAC} that
$
\{\hat W_i = W_i\}  = \{\hat U_i^n = U_i^n\}
$
and
$
\{\hat W_{[i-1]} = W_{[i-1]}\} = \{\hat X_{[i-1]}^n = X_{[i-1]}^n\}$.
For each $i\in\mathcal{I}$ and each $k\in \mathcal{J}_i$, we have
\begin{align}
&\Pr\big\{\{\hat U_{i,k} \ne U_{i,k}\}\cap \{\hat U_i^{k-1}=U_i^{k-1}\} \cap \{\hat X_{[i-1]}^n = X_{[i-1]}^n\}\big\} \notag\\*
& \le \sum_{u_{i,k}\in\{0,1\}}\sum_{u_i^{k-1}\in\mathcal{U}_i^{k-1}}\sum_{x_{[i-1]}^n\in \mathcal{X}_{[i-1]}^n}\int_{\mathcal{Y}^n} p_{U_i^{k}, X_{[i-1]}^n, Y^n}(u_i^{k},x_{[i-1]}^n, y^n)  \notag\\*
&\qquad \times \mathbf{1}\left\{p_{U_{i,k}|U_i^{k-1}, X_{[i-1]}^n, Y^n}(u_{i,k}|u_i^{k-1}, x_{[i-1]}^n, y^n)\le p_{U_{i,k}|U_i^{k-1}, X_{[i-1]}^n, Y^n}(1-u_{i,k}|u_i^{k-1}, x_{[i-1]}^n, y^n)\right\} \mathrm{d}y^n\label{eq2aPropositionErrorPolarMAC} \\
&\le 2\sum_{u_i^{k-1}\in\mathcal{U}_i^{k-1}}\sum_{x_{[i-1]}^n\in \mathcal{X}_{[i-1]}^n}\int_{\mathcal{Y}^n} p_{U_i^{k-1}, X_{[i-1]}^n, Y^n}(u_i^{k-1},x_{[i-1]}^n, y^n)\notag\\*
&\qquad \times \sqrt{p_{U_{i,k}|U_i^{k-1}, X_{[i-1]}^n, Y^n}(0|u_i^{k-1}, x_{[i-1]}^n, y^n)p_{U_{i,k}|U_i^{k-1}, X_{[i-1]}^n, Y^n}(1|u_i^{k-1}, x_{[i-1]}^n, y^n)}\mathrm{d}y^n \\*
& =  Z^{[p_{X_\mathcal{I}}; q_{Y|X_\mathcal{I}}]}(U_{i,k}|U_i^{k-1}, X_{[i-1]}^n, Y^n) \label{eq2PropositionErrorPolarMAC}
\end{align}
where \eqref{eq2aPropositionErrorPolarMAC} follows from~\eqref{defSCdecoderMAC}.
In addition, it follows from~\eqref{frozenBitsMAC} and~\eqref{defSCdecoderMAC} that
\begin{align}
\Pr\big\{\{\hat U_{i,k} \ne U_{i,k}\}\cap \{\hat U_i^{k-1}=U_i^{k-1}\} \cap \{\hat X_{[i-1]}^n = X_{[i-1]}^n\}\big\}=0 \label{eq3PropositionErrorPolarMAC}
\end{align}
for each $i\in\mathcal{I}$ and each $k\in \mathcal{J}_i^c$. Combining~\eqref{eq1PropositionErrorPolarMAC}, \eqref{eq2PropositionErrorPolarMAC} and~\eqref{eq3PropositionErrorPolarMAC}, we obtain~\eqref{st1PropErrorPolarMAC}.

\section{Proof of Lemma~\ref{lemmaGoodA}}\label{appendixB}
%\begin{IEEEproof}[] %\label{appendixA}
Let $q_{Y|X}$ be the conditional distribution that characterizes the AWGN channel and fix a $\gamma\in[0, 1)$.
 %We will show the existence of $\mathcal{A}$ that satisfies~\eqref{lemmaGoodAst1}, \eqref{lemmaGoodAst2} and~\eqref{lemmaGoodAst3} by a probabilistic argument. To this end, define a random set
%\begin{equation}
%\mathfrak{A}\triangleq \{A_1, A_2, \ldots, A_n\} \label{defSetA}
%\end{equation}
%consisting of~$n$ i.i.d.\ random variables where $A_1\sim \mathcal{N}(a_1; 0, P-\frac{1}{n^{1/3}})$.
Recall the definitions of $p_X^\prime$ and $s_X$ in~\eqref{defDistPprime} and \eqref{defDistS} respectively and recall that $\Phi_X$ is the cdf of $s_X$.
Fix a sufficiently large~$n \ge 36$ that satisfies
\begin{equation}
\frac{1}{n^{(1-\gamma)/\beta}} \in \left(0,\frac{1+P}{2}\right] \label{sufficientlyLargeNappendixA*}
\end{equation}
and
\begin{equation}
\Phi_X^{-1} \left( \frac{1}{n^{1-(1-\gamma)/\beta}}\right) \le -1. \label{sufficientlyLargeNappendixA}
\end{equation}
%
%
%
%let $Z$ be a Gaussian random variable whose mean and variance are~$0$ and $P - \frac{1}{n^{1/3}}$ respectively, and let $\hat Z$ be a quantized version of~$Z$ defined as follows:
In addition, recall the definition of~$\mathcal{X}$ in~\eqref{defSetX} and let $g:\mathbb{R}\rightarrow\mathcal{X}$ be a quantization function such that
\begin{equation}
g(t)\triangleq \Phi_X^{-1}\left(\frac{\ell}{n}\right) \label{defFunctionG}
\end{equation}
where $\ell\in\{1, 2, \ldots,\frac{n}{2}, \ldots n-1\}$ is the unique integer that satisfies
\begin{align}
\begin{cases}
\Phi_X^{-1}\left(\frac{\ell}{n}\right) \le t < \Phi_X^{-1}\left(\frac{\ell+1}{n} \right) & \text{if $t\ge 0$,} \\
\Phi_X^{-1}\left(\frac{\ell-1}{n}\right) < t \le \Phi_X^{-1}\left(\frac{\ell}{n} \right) & \text{if $t< 0$.}
\end{cases}
\end{align}
In words, $g$ quantizes every $a\in\mathbb{R}$ to its nearest point in $\mathcal{X}$ whose magnitude is smaller than~$|a|$.
Let
\begin{equation}
\hat X \triangleq g(X) \label{defHatX}
\end{equation}
be the quantized version of~$X$. By construction,
\begin{equation}
\Pr_{s_X}\left\{|\hat X| \le |X|\right\}=1, \label{byConstruction0}
\end{equation}
\begin{equation}
\Pr_{s_X}\left\{\left|\Phi_X(\hat X) - \Phi_X(X)\right| \le \frac{1}{n}\right\}=1 \label{byConstruction1}
\end{equation}
and
\begin{equation}
\Pr_{s_X}\left\{X\in \left(\Phi_X^{-1}\left(\frac{\ell}{n}\right), \Phi_X^{-1}\left(\frac{\ell+1}{n}\right)\right)\right\}=\frac{1}{n} \label{byConstruction2}
\end{equation}
for all $\ell\in\{0, 1, \ldots, n-1\}$.
It follows from~\eqref{byConstruction2} and the definition of $p_X^\prime$ in~\eqref{defDistPprime} that
\begin{align}
\E_{p_X^\prime}[X^2]=\E_{s_X}[\hat X^2] \label{lemmaGoodAst1proof*}
\end{align}
and
\begin{align}
\Pr_{p_{X^n}^\prime}\left\{\frac{1}{n}\sum_{k=1}^n X_k^2 >P\right\}=\Pr_{s_{X^n}}\left\{\frac{1}{n}\sum_{k=1}^n \hat X_k^2>P\right\} \label{lemmaGoodAst2proof*}
\end{align}
where $s_{X^n}(x^n)\triangleq \prod_{k=1}^n s_X(x_k)$.
Consequently, in order to show~\eqref{lemmaGoodAst1} and \eqref{lemmaGoodAst2}, it suffices to show
\begin{align}
\E_{s_X}[\hat X^2]\le P\left(1-\frac{1}{n^{(1-\gamma)/\beta}}\right) \label{lemmaGoodAproofSt1}
\end{align}
and
\begin{align}
\Pr_{s_{X^n}}\left\{\frac{1}{n}\sum_{k=1}^n \hat X_k^2>P\right\}\le  \frac{e^3}{e^{n^{\frac{1}{2}-\frac{1-\gamma}{\beta}}}} \label{lemmaGoodAproofSt2}
\end{align}
respectively. Using~\eqref{byConstruction0} and the definition of $s_X$ in~\eqref{defDistS}, we obtain~\eqref{lemmaGoodAproofSt1}. In order to show~\eqref{lemmaGoodAproofSt2}, we consider the following chain of inequalities:
%Using \eqref{st3thmPolarAWGNavg}, \eqref{st4thmPolarAWGNavg} and the Markov's inequality, we have
\begin{align}
&\Pr_{s_{X^n}}\left\{\frac{1}{n}\sum_{k=1}^n \hat X_k^2>P\right\} \notag\\
& \quad\le \Pr_{s_{X^n}}\left\{\frac{1}{n}\sum_{k=1}^n X_k^2>P\right\}  \label{lemmaGoodAproofSt2*a}\\
& \quad = \Pr_{s_{X^n}}\left\{\frac{\sum_{k=1}^n X_k^2}{\sqrt{n}P\big(1-\frac{1}{n^{(1-\gamma)/\beta}}\big)}> \sqrt{n} + \frac{n^{\frac{1}{2}-\frac{1-\gamma}{\beta}}}{1-\frac{1}{n^{(1-\gamma)/\beta}}}\right\}  \label{lemmaGoodAproofSt2*b}\\
&\quad \le \Pr_{s_{X^n}}\Bigg\{e^{\frac{\sum_{k=1}^n X_k^2}{\sqrt{n}P\big(1-\frac{1}{n^{(1-\gamma)/\beta}}\big)}}> e^{\left(\sqrt{n}+ n^{\frac{1}{2}-\frac{1-\gamma}{\beta}}\right)}\Bigg\}  \label{lemmaGoodAproofSt2*c}\\
&\quad \le \left(1-\frac{1}{\sqrt{n}/2}\right)^{-n/2} e^{-\left(\sqrt{n} + n^{\frac{1}{2}-\frac{1-\gamma}{\beta}}\right)}\label{lemmaGoodAproofSt2*d}\\
&\quad = \left(1+\frac{1}{\sqrt{n}/2 -1}\right)^{n/2} e^{-\left(\sqrt{n} + n^{\frac{1}{2}-\frac{1-\gamma}{\beta}}\right)}\\
&\quad \le e^{\frac{n}{\sqrt{n}-2}} \cdot e^{-\left(\sqrt{n} + n^{\frac{1}{2}-\frac{1-\gamma}{\beta}}\right)}\label{lemmaGoodAproofSt2*e}\\
& \quad = e^{\frac{2\sqrt{n}}{\sqrt{n}-2}}\cdot e^{- n^{\frac{1}{2}-\frac{1-\gamma}{\beta}}} \\
&\quad \le e^3 \cdot e^{- n^{\frac{1}{2}-\frac{1-\gamma}{\beta}}}\label{lemmaGoodAproofSt2*}
\end{align}
where
\begin{itemize}
\item \eqref{lemmaGoodAproofSt2*a} is due to~\eqref{byConstruction0}.
\item \eqref{lemmaGoodAproofSt2*d} is due to Markov's inequality.
\item \eqref{lemmaGoodAproofSt2*e} is due to the fact that $(1+\frac{1}{t})^t \le e$ for all $t>0$.
\item \eqref{lemmaGoodAproofSt2*} is due to the assumption that $\sqrt{n}\ge 6$.
\end{itemize}
It remains to show~\eqref{lemmaGoodAst3}. To this end, we let $q_Z$ denote the distribution of the standard normal random variable (cf.\ \eqref{channelLaw}) and consider
% ~\eqref{lemmaGoodAst3}
%and
%\begin{align}
%\mathrm{C}(P)- I_{p_X^\prime q_Z}(X;X+Z) \le \frac{1}{n^{1/3}}\label{lemmaGoodAproofSt3}
%\end{align}
%%respectively and
%%\begin{align}
%% = I_{s_X q_Z}(\hat X;\hat X+Z) \label{lemmaGoodAst3proof*}
%%\end{align}
%respectively where .
\begin{align}
&\mathrm{C}\left(P-\frac{1}{n^{(1-\gamma)/\beta}}\right) - I_{p_X^\prime q_{Y|X}}(X;Y)\notag\\*
&\quad =\mathrm{C}\left(P-\frac{1}{n^{(1-\gamma)/\beta}}\right) - I_{p_X^\prime q_Z}( X;X+Z) \\*
%&\quad = I_{s_{X} q_{Z}}(X;X+Z) - I_{s_X q_Z}(\hat X;\hat X+Z) \label{lemmaGoodAproofEq4a} \\*
&\quad = I_{s_{X} q_{Z}}(X;X+Z) - I_{p_X^\prime q_Z}( X;X+Z)
%& \quad = h_{s_{X} q_{Z}}(X+Z) - h_{s_{X} q_{Z}}(\hat X + Z)\\
%&\quad =
\label{lemmaGoodAproofSt3}
\end{align}
where \eqref{lemmaGoodAproofSt3} is due to the definition of $s_X$ in~\eqref{defDistS}.
In order to simplify the RHS of~\eqref{lemmaGoodAproofSt3}, we invoke~\cite[Corollary~4]{PolyanskiyWu16} and obtain
\begin{align}
I_{s_{X} q_{Z}}(X;X+Z) - I_{p_X^\prime q_Z}( X;X+Z) \le (\log e)(3\sqrt{1+P}+4\sqrt{P})W_2(s_{Y}, p_Y^\prime). \label{lemmaGoodAproofSt4}
\end{align}
After some tedious calculations which will be elaborated after this proof, it can be shown that the Wasserstein distance in~\eqref{lemmaGoodAproofSt4} satisfies
\begin{equation}
W_2(s_{Y}, p_Y^\prime) \le \sqrt{\frac{\kappa\log n}{n^{\frac{2(1-\gamma)}{\beta}}}}\:, \label{lemmaInAppendix}
\end{equation}
where
\begin{equation}
\kappa\triangleq P^2 + 4P+ \frac{4P}{\log e}\left(1 + \log\sqrt{\frac{P}{2\pi}}\right). \label{defKappa}
\end{equation}
%whose derivation will be given after this proof.
%
%\begin{align}
%&\E_{p_{A^n}}\left[(p_Y^*(y))^2\right] \notag\\
%&\quad = \frac{1}{n^2}\left(\sum_{k=1}^n \E_{p_{A^n}}\!\!\left[(q_{Y|X}(y|A_k))^2\right] + 2\sum_{\ell<k}\E_{p_{A^n}}\!\!\left[q_{Y|X}(y|A_\ell)\right]\E_{p_{A^n}}\!\!\left[q_{Y|X}(y|A_k)\right]\right)\\
%&\quad =\frac{1}{n}\left(\E_{p_{A^n}}\!\!\left[(q_{Y|X}(y|A_1))^2\right] + n(n-1)(s_Y(y))^2\right),
%\end{align}
%it follows from~\eqref{lemmaGoodAproofEq5b} that
%\begin{align}
%&\mathrm{C}\left(P-\frac{1}{n^{1/3}}\right) - \E_{p_{A^n}}\left[I_{p_{X}^* q_{Y|X}}(X;Y)\right] \notag\\
%&\quad \le \frac{1}{n}\left( \int_{\mathcal{Y}} \frac{\E_{p_{A^n}}\!\!\left[(q_{Y|X}(y|A_1))^2\right]}{s_Y(y)}\,\mathrm{d}y -1\right)\\
%&\quad = \frac{P-\frac{1}{n^{1/3}}}{n}. \label{lemmaGoodAproofEq6}
%\end{align}
On the other hand, since
\begin{equation}
\frac{\log(1+P-\xi)}{\log e} \ge \frac{\log (1+P)}{\log e} - \frac{\xi}{1+P} - \frac{2\xi^2}{(1+P)^2}
\end{equation}
for each $\xi \in \big(0,\frac{1+P}{2}\big]$ by Taylor's theorem and $\frac{1}{n^{(1-\gamma)/\beta}} \in \big(0,\frac{1+P}{2}\big]$ by~\eqref{sufficientlyLargeNappendixA*}, we have
\begin{align}
\mathrm{C}\left(P-\frac{1}{n^{\frac{1-\gamma}{\beta}}}\right) \ge \mathrm{C}(P) -\frac{\log e}{2} \left(\frac{1}{n^{\frac{1-\gamma}{\beta}}(1+P)} + \frac{2}{n^{\frac{2(1-\gamma)}{\beta}}(1+P)^2}\right). \label{lemmaGoodAproofEq7}
\end{align}
Using \eqref{lemmaGoodAproofSt3}, \eqref{lemmaGoodAproofSt4}, \eqref{lemmaInAppendix} and~\eqref{lemmaGoodAproofEq7}, we obtain
\begin{equation}
\mathrm{C}\left(P\right) - I_{p_X^\prime q_{Y|X}}(X;Y) \le  (\log e)(3\sqrt{1+P}+4\sqrt{P})\sqrt{\frac{\kappa\log n}{n^{2(1-\gamma)/\beta}}} +\frac{\log e}{2} \left(\frac{1}{n^{\frac{1-\gamma}{\beta}}(1+P)} + \frac{2}{n^{\frac{2(1-\gamma)}{\beta}}(1+P)^2}\right).
\end{equation}
Consequently, \eqref{lemmaGoodAst3} holds for some constant $t^\prime>0$ that does not depend on~$n$.
%\end{IEEEproof}
%\begin{align}
%\frac{1}{n}\sum_{k=1}^n a_k^2 \le P - \frac{1}{n^{1/3}}, \label{lemmaGoodAProofSt1}
%\end{align}
%\begin{align}
%\frac{1}{n}\sum_{k=1}^n a_k^4 \le 3P^2 \label{lemmaGoodAProofSt2}
%\end{align}

%\section{Proof of Lemma~\ref{lemmaVD}} \label{appendixB}
\section*{Derivation of~\eqref{lemmaInAppendix}}
%Using Corollary~4 in~\cite{PolyanskiyWu16}, we obtain
%\begin{align}
%I_{s_{X} q_{Z}}(X;X+Z) - I_{p_X^\prime q_Z}( X;X+Z) \le (\log e)(3\sqrt{1+P}+4\sqrt{P})W_2(s_{Y}, p_Y^\prime) \label{lemmaGoodAproofSt4}
%\end{align}
%where $p_{Y}^\prime$ and $s_Y$ denote the marginal distributions of $p_X^\prime q_{Y|X}$ and $s_Xq_{Y|X}$ respectively.
Consider the distribution (coupling) $r_{X, \hat X, Y, \hat Y}$ defined as
 \begin{equation}
 r_{X, \hat X, Y, \hat Y}(x, \hat x, y, \hat y)=s_{X}(x)q_Z(y-x)\mathbf{1}\{\hat x=g(x)\}\mathbf{1}\{\hat y=g(x)+y-x\} \label{defDistRjoint}
 \end{equation}
and simplify the Wasserstein distance in~\eqref{lemmaGoodAproofSt4} as follows:
\begin{align}
\left(W_2(s_{Y}, p_Y^\prime) \right)^2
& \le \int_{\mathbb{R}}\int_{\mathbb{R}} r_{Y, \hat Y}(y, \hat y) (y-\hat y)^2  \mathrm{d}\hat y\mathrm{d}y \label{lemmaGoodAproofSt5a} \\*
%& = \int_{\mathbb{R}}\int_{\mathbb{R}}\int_{\mathbb{R}} s_{X, \hat X, X+Z, \hat X+Z}(x, \hat x, x+z, \hat x+z) (x-\hat x)^2  \mathrm{d}z \mathrm{d}\hat x \mathrm{d}x \\
& = \int_{\mathbb{R}}\int_{\mathbb{R}}r_{X, \hat X}(x, \hat x) (x-\hat x)^2  \mathrm{d}\hat x \mathrm{d}x \label{lemmaGoodAproofSt5b} \\*
& = \int_{\mathbb{R}}s_X(x)(x-g(x))^2 \mathrm{d}x  \label{lemmaGoodAproofSt5}
\end{align}
where
\begin{itemize}
\item \eqref{lemmaGoodAproofSt5a} follows from the definition of $W_2$ in~\eqref{defW2Distance} and the facts due to~\eqref{defDistRjoint} that $r_{Y}=s_Y$ and $r_{\hat Y}=p_Y^\prime$.
    \item  \eqref{lemmaGoodAproofSt5b} follows from the fact due to~\eqref{defDistRjoint} that $\Pr_{r_{X, \hat X, Y, \hat Y}}\{Y-\hat Y = X-\hat X\}=1$.
\item \eqref{lemmaGoodAproofSt5} is due to~\eqref{defDistRjoint}.
\end{itemize}
Following~\eqref{lemmaGoodAproofSt5}, we define $\xi_n$ to be the positive number that satisfies
\begin{align}
\int_{\xi_n}^\infty s_X(x) \mathrm{d}x  = \frac{1}{n^{1-(1-\gamma)/\beta}} \label{defXiN}
\end{align}
and consider
\begin{align}
&\int_{\mathbb{R}}s_X(x)(x-g(x))^2 \mathrm{d}x \notag\\*
&\quad = \int_{-\infty}^{-\xi_n}s_X(x)(x-g(x))^2 \mathrm{d}x + \int_{-\xi_n}^{\xi_n}s_X(x)(x-g(x))^2 \mathrm{d}x + \int_{\xi_n}^{\infty}s_X(x)(x-g(x))^2 \mathrm{d}x \\
&\quad \le \int_{-\infty}^{-\xi_n}s_X(x)x^2 \mathrm{d}x + \int_{-\xi_n}^{\xi_n}s_X(x)(x-g(x))^2 \mathrm{d}x + \int_{\xi_n}^{\infty}s_X(x)x^2 \mathrm{d}x \label{lemmaGoodAproofSt6a}\\
&\quad = 2\int_{\xi_n}^{\infty}s_X(x)x^2 \mathrm{d}x +  \int_{-\xi_n}^{\xi_n}s_X(x)(x-g(x))^2 \mathrm{d}x \label{lemmaGoodAproofSt6}
%&\quad = \int_{\mathbb{R}}\frac{1}{\sqrt{2\pi\left(P-\frac{1}{n^{1/3}}\right)}}e^{-\frac{x^2}{2\left(P-\frac{1}{n^{1/3}}\right)}}(x-g(x))^2 \mathrm{d}x
\end{align}
where~\eqref{lemmaGoodAproofSt6a} follows from the fact due to~\eqref{defFunctionG} that $x-g(x) \ge 0$ for all $x\in\mathbb{R}$. In order to bound the first term in~\eqref{lemmaGoodAproofSt6}, we let
\begin{equation}
P_n\triangleq P\Big(1-\frac{1}{n^{(1-\gamma)/\beta}}\Big)
\end{equation}
and consider
\begin{align}
\int_{\xi_n}^{\infty}s_X(x)x^2 \mathrm{d}x %&= \sqrt{\frac{P-\frac{1}{n^{1/3}}}{2\pi}}\left( \sqrt{2\pi\left(P-\frac{1}{n^{1/3}}\right)}\int_{\xi_n}^\infty s_X(x) \mathrm{d}x +\xi_n e^{-\frac{\xi_n^2}{2\left(P-\frac{1}{n^{1/3}}\right)}} \right)\\
&= P_n\left(\int_{\xi_n}^{\infty}s_X(x) \mathrm{d}x + \xi_n s_X(\xi_n)\right) \label{lemmaGoodAproofSt6*a} \\*
&< \int_{\xi_n}^{\infty}s_X(x) \mathrm{d}x \left(2P_n+\xi_n^2\right) \label{lemmaGoodAproofSt6*b}\\*
& < \frac{1}{n^{1-(1-\gamma)/\beta}} \left(2P+\xi_n^2\right)\label{lemmaGoodAproofSt6*}
\end{align}
where
\begin{itemize}
\item \eqref{lemmaGoodAproofSt6*a} follows from integration by parts.
\item  \eqref{lemmaGoodAproofSt6*b} is due to the simple fact that
\begin{align}
\int_{\xi_n}^{\infty} s_X(x) \mathrm{d}x &> \frac{\xi_n^2}{P_n+\xi_n^2}\int_{\xi_n}^{\infty} \left(1+\frac{P_n}{x^2}\right)s_X(x) \mathrm{d}x \\*
&= \left(\frac{P_n}{P_n+\xi_n^2}\right)\xi_ns_X(\xi_n).
\end{align}
\item \eqref{lemmaGoodAproofSt6*} involves using~\eqref{defXiN}.
\end{itemize}
In order to bound the term in~\eqref{lemmaGoodAproofSt6*},
% In order to bound the second term in~\eqref{lemmaGoodAproofSt6},
we note that
\begin{equation}
\xi_n \ge 1 \label{eqnXi>1}
\end{equation}
by~\eqref{sufficientlyLargeNappendixA} and~\eqref{defXiN} and would like to obtain an upper bound on $\xi_n$ through the following
chain of inequalities:
\begin{align}
\frac{1}{n^{1-(1-\gamma)/\beta}}%&\le\frac{1}{n^{2/3}} \\
 &=\int_{\xi_n}^\infty s_X(x) \mathrm{d}x \label{lemmaGoodAproofSt6**a}\\
&\le \int_{\xi_n}^\infty x s_X(x) \mathrm{d}x \label{lemmaGoodAproofSt6**b}\\
&=  P_n s_X(\xi_n) \label{lemmaGoodAproofSt6**c} \\
& = \sqrt{\frac{P_n}{2\pi}}\, e^{-\frac{\xi_n^2}{2 P_n}}\\
& < \sqrt{\frac{P}{2\pi}}\, e^{-\frac{\xi_n^2}{2 P}}
\label{lemmaGoodAproofSt6**}
\end{align}
where
\begin{itemize}
\item \eqref{lemmaGoodAproofSt6**a} is due to~\eqref{defXiN}.
\item \eqref{lemmaGoodAproofSt6**b} is due to~\eqref{eqnXi>1}.
\end{itemize}
Since
\begin{equation}
\xi_n^2 \le \frac{2P}{\log e}\left(\bigg(1-\frac{1-\gamma}{\beta}\bigg)\log n + \log\sqrt{\frac{P}{2\pi}}\right)
\end{equation}
by~\eqref{lemmaGoodAproofSt6**} and $\beta=4.714>3$, it follows from~\eqref{lemmaGoodAproofSt6*} that
\begin{align}
\int_{\xi_n}^{\infty}s_X(x)x^2 \mathrm{d}x &<  \frac{1}{n^{1-\frac{1-\gamma}{\beta}}} \left(2P+ \frac{2P}{\log e}\left(\bigg(1-\frac{1-\gamma}{\beta}\bigg)\log n + \log\sqrt{\frac{P}{2\pi}}\right)\right)\\*
%&<  \frac{\log n}{n^{1-\frac{1-\gamma}{\beta}}} \left(2P+ \frac{2P}{\log e}\left(1+ \log\sqrt{\frac{P}{2\pi}}\right)\right) \\*
& < \frac{\log n}{n^{\frac{2(1-\gamma)}{\beta}}} \left(2P+ \frac{2P}{\log e}\left(1+ \log\sqrt{\frac{P}{2\pi}}\right)\right). \label{lemmaGoodAproofSt6****}
\end{align}
%To simpify notation, define
%\begin{equation}
%\xi_n\triangleq \sqrt{\frac{P}{\log e} \log \left(\frac{Pn}{2\pi}\right)}\,, \label{defCn}
%\end{equation}
%and it follows from \eqref{lemmaGoodAproofSt6**} that
%\begin{equation}
%\xi_n\le \xi_n. \label{eqnXnLessThanCn}
%\end{equation}
In order to bound the second term in~\eqref{lemmaGoodAproofSt6}, we consider
\begin{align}
& \int_{-\xi_n}^{\xi_n}s_X(x)(x-g(x))^2 \mathrm{d}x \notag\\
%&\quad \le \int_{-\xi_n}^{\xi_n}s_X(x)(x-g(x))^2 \mathrm{d}x   \label{lemmaGoodAproofSt6***a}\\
 &\quad = \int_{-\xi_n}^{\xi_n}s_X(x)\Big(\Phi^{-1}\big(\Phi(x)\big)-\Phi^{-1}\big(\Phi(g(x))\big)\Big)^2 \mathrm{d}x  \\
 &\quad \le  \int_{-\xi_n}^{\xi_n}s_X(x)\left(\frac{1}{n}\times \frac{1}{s_X(\xi_n)}\right)^2 \mathrm{d}x  \label{lemmaGoodAproofSt6***b} \\
 &\quad \le  \int_{-\xi_n}^{\xi_n}s_X(x)\left(\frac{P_n^2}{n^{\frac{2(1-\gamma)}{\beta}}}\right) \mathrm{d}x \label{lemmaGoodAproofSt6***c} \\*
 &\quad < \frac{P^2}{n^{\frac{2(1-\gamma)}{\beta}}} \label{lemmaGoodAproofSt6***}
\end{align}
where
\begin{itemize}
%\item \eqref{lemmaGoodAproofSt6***a} is due to~\eqref{eqnXnLessThanCn}.
\item  \eqref{lemmaGoodAproofSt6***b} is due to~\eqref{byConstruction1}, the mean value theorem and the fact that the derivative of $\Phi$ is always positive and uniformly bounded below by $s_X(\xi_n)$ on the interval $[-\xi_n, \xi_n]$.
    \item \eqref{lemmaGoodAproofSt6***c} is due to~\eqref{lemmaGoodAproofSt6**c}.
\end{itemize}
Combining~\eqref{lemmaGoodAproofSt5}, \eqref{lemmaGoodAproofSt6},  \eqref{lemmaGoodAproofSt6****} and~\eqref{lemmaGoodAproofSt6***} and recalling the definition of~$\kappa$ in~\eqref{defKappa},
we obtain~\eqref{lemmaInAppendix}.
\end{document}